\documentclass[pra,twocolumn,showpacs,floatfix]{revtex4}
\usepackage{graphicx}
\usepackage{amsmath}

\begin{document}

\title{Few-body spin couplings and their implications for universal quantum computation}
\author{Ryan Woodworth$^1$, Ari Mizel$^{1,2}$, and Daniel A. Lidar$^3$}
\affiliation{$^1$Physics Department and $^2$Materials Research Institute, Pennsylvania
State University, University Park, PA 16802}
\affiliation{$^{3}$Chemical Physics Theory Group, and Center for Quantum Information and
Quantum Control, University of Toronto, 80 St. George St., Toronto, Ontario
M5S 3H6, Canada}

\begin{abstract}
Electron spins in semiconductor quantum dots are promising candidates for the experimental realization of solid-state qubits. We analyze the dynamics of a system of three qubits arranged in a linear geometry and a system of four qubits arranged in a square geometry. Calculations are performed for several quantum dot confining potentials. In the three-qubit case, three-body effects are identified that have an important quantitative influence upon quantum computation. In the four-qubit case, the full Hamiltonian is found to include both three-body and four-body interactions that significantly influence the dynamics in physically relevant parameter regimes. We consider the implications of these results for the encoded universality paradigm applied to the four-electron qubit code; in particular, we consider what is required to circumvent the four-body effects in an encoded system (four spins per encoded qubit) by the appropriate tuning of experimental parameters.
\end{abstract}

\maketitle

\section{Introduction}

Electron spins in semiconductor quantum dots are a leading candidate for the
physical realization of qubits in a quantum computer \cite{LossDiVincenzo98}. Although any quantum algorithm can be implemented using single-qubit and
two-qubit gates \cite{MAN/ILC}, many such algorithms realize substantial
increases in efficiency by exploiting simultaneous interactions among three
or more qubits \cite
{Bacon:99a,Kempe:00,DiVincenzo:00a,LidarWu:01,Bacon:Sydney,Freedman:02,Raussendorf:01,Lidar:00b,Steane:99a,Preskill,Gottesman:97a,Shor:96,Farhi:01,Bacon:01}. In order to employ such simultaneous interactions, it is essential to
understand in detail the many-body dynamics of the system of coupled qubits.
More generally, since a practical quantum computer may need to contain as
many as $10^{6}$ qubits \cite{Preskill}, it is essential to characterize the
effect of many-body interactions on the system's overall energy landscape.

In past work \cite{MizelLidar,MizelLidarB}, we used a model confining potential of
superposed parabolic minima to demonstrate that three-body effects
significantly influence the Hamiltonian of three electrons confined to three
quantum dots at the vertices of an equilateral triangle and that four-body
effects are significant for four electrons confined to a tetrahedral
arrangement of four dots. Here we extend these results in two ways. First,
we analyze three quantum dots in a linear geometry \cite{3inaline_figure}
and four dots in a square geometry \cite{Burkard:00} since these geometries
are more likely to occur in a real quantum computer apparatus. Second, by
employing a Gaussian shape for the confining potential of each well \cite{Hu:99}, we explore the sensitivity of the many-body effects to the form of
the confining potential. In both cases, a non-perturbative calculation finds
that many-body effects contribute appreciably to the Hamiltonian. We
note that Scarola et al. \cite{Scarola:04,Scarola:05} have demonstrated that the application of a
magnetic field allows chiral terms to arise in the spin Hamiltonian, which
modifies this Hamiltonian in another important manner as compared to the
naive Heisenberg form.

To date, discussions of quantum dot quantum computation have nearly always
assumed pairwise Heisenberg interactions. In view of the above result, this
implies that computational errors may occur in the context of quantum
computers using electron spin qubits in quantum dots, unless one always
simultaneously couples only disjoint pairs of dots. There are at least
four circumstances where this may be undesirable or
even infeasible. One is fault tolerant quantum error correction, where
simultaneous operations on several coupled dots have been associated with
better error thresholds. A second is adiabatic
quantum computation \cite{Farhi:01}, in which the final Hamiltonian
may include the simultaneous
interactions that we discuss here. We will not analyze these possibilities
here, although we believe that the methods we discuss below are relevant to
them.

We will focus on two other contexts, that of \textquotedblleft encoded
universality\textquotedblright\ (EU) \cite{Bacon:99a,Kempe:00,DiVincenzo:00a,Bacon:Sydney,LidarWu:01} and that of
computation on decoherence-free subspaces (DFSs) \cite{Bacon:99a,Kempe:00,Zanardi:99a} and supercoherent qubits \cite{Bacon:01}.
In these cases, the goal is to perform universal quantum computation using
(EU:) only the most easily controllable interaction, or (DFS,
supercoherence:) using only interactions that preserve the code subspace,
since that subspace offers protection against certain types of decoherence.
(Strong and fast exchange interaction pulses can further be used to suppress
decoherence \cite{WuLidar:01b} and to eliminate decoherence-induced leakage 
\cite{WuByrdLidar:02}.)

We will refer to these cases collectively as \textquotedblleft encoded
quantum computation.\textquotedblright \quad It turns out that universal quantum
computation using only the Heisenberg exchange interaction is an extremely
attractive possibility in encoded models, and we will consider it in detail
below. After establishing that four-body interaction terms can arise in a
Heisenberg exchange Hamiltonian, we investigate the question of neutralizing
their effect by using encoded qubits \cite{Zanardi:97c,Bacon:99a,Kempe:00,Bacon:Sydney,LidarWu:01,Zanardi:99a,WuLidar:01b,WuByrdLidar:02,Lidar:PRA00Exchange,Bacon:01,Lidar:DFSreview}. By generalizing the work of Bacon \cite{Bacon:thesis}, who showed that
universal quantum computation was possible using encoded gates with two-body
coupling Hamiltonians (i.e., assuming that the Heisenberg
Hamiltonian was applicable even when coupling three or more dots at a time), we enumerate tuning conditions on experimental parameters that are needed
for the four-body effects to cancel out. An alternative is to design
these encoded gates while allowing only pairs of electrons to couple at any
given time. This is indeed possible, as shown in Ref. \cite{Hsieh:04}, for the price of significantly longer pulse
sequences per given encoded gate. Nevertheless, in view of the findings
reported here and in Refs. \cite{Scarola:04,Scarola:05}, this price may be
worth paying.

\section{Three-Electron Case}

A system of three electrons within a confining scalar potential $V(\mathbf{r})$ obeys the Hamiltonian 
\begin{eqnarray}
H &=&\sum_{i=1}^{3}\left[ \frac{\mathbf{p}_{i}^{2}}{2m}+V(\mathbf{r}_{i})\right] +\sum_{i<j}\frac{e^{2}}{\kappa \mid \mathbf{r}_{i}-\mathbf{r}_{j}\mid }  \label{Hcoord_3} \\
&\equiv &\sum_{i=1}^{3}h(\mathbf{r}_{i})+\sum_{i<j}w(\mathbf{r}_{i},\mathbf{r}_{j})  \label{H_hw_3}
\end{eqnarray}
in the absence of spin-orbit coupling and external magnetic fields.
Although Ref. \cite{MizelLidar} succeeded in demonstrating significant three-body and four-body effects in systems containing three or more electrons, a confining potential with quadratic minima has certain other characteristics which are unlikely
to describe an experimental arrangement; for example, it diverges at large
distances from the origin, and the single adjustable parameter $\omega_o$
forces us to specify very narrow minima whenever we want a high barrier
between them. We therefore begin with the Gaussian form
\begin{equation}
V(\mathbf{r})=-V_{0}[e^{-\alpha \mid \mathbf{r}-\mathbf{A}\mid
^{2}}+e^{-\alpha \mid \mathbf{r}-\mathbf{B}\mid ^{2}}+e^{-\alpha \mid 
\mathbf{r}-\mathbf{C}\mid ^{2}}],  \label{V_Exp_3}
\end{equation}
which has two tunable parameters.  The three fixed points are collinear and separated by a distance $2 l$: $\mathbf{A} = (-2 l,0,0)$, $\mathbf{B} = (0,0,0)$, and $\mathbf{C} = (2 l,0,0)$.

We assume a
Heitler-London approximation \cite{HL}, wherein excited orbital states and
states with double occupation of any single dot are neglected (see
Ref. \cite{Scarola:05} for a recent discussion of the validity of this
approximation in the context of electron spin qubits). The system's only
degrees of freedom are therefore the spins of the confined electrons,
leading to a total of $2^{3}=8$ \textquotedblleft
computational\textquotedblright\ basis states 
\begin{equation}
\left\vert \Psi (s_{A},s_{B},s_{C})\right\rangle =\sum_{P}\delta
_{P}P[\left\vert A\right\rangle \left\vert B\right\rangle \left\vert
C\right\rangle \left\vert s_{A}\right\rangle \left\vert s_{B}\right\rangle
\left\vert s_{C}\right\rangle ].  \label{psicomp_3}
\end{equation}
In the above, $\left\vert \{A\}\right\rangle $ are the three localized
orbital ground states; $\left\vert s_{\{A\}}\right\rangle $ denote the
corresponding spin states; $P$ is the set of all permutations of $\{A,B,C\}$; and $\delta _{P}$ is 1 (-1) for even (odd) permutations. For instance, one
of the eight (unnormalized) basis states is 
\begin{eqnarray*}
|\Psi (\uparrow \uparrow \downarrow )\rangle  &=&|ABC\rangle \mid \uparrow
\uparrow \downarrow \rangle -|ACB\rangle \mid \uparrow \downarrow \uparrow
\rangle  \\
&+&|CAB\rangle \mid \downarrow \uparrow \uparrow \rangle -|CBA\rangle \mid
\downarrow \uparrow \uparrow \rangle  \\
&+&|BCA\rangle \mid \uparrow \downarrow \uparrow \rangle -|BAC\rangle \mid
\uparrow \uparrow \downarrow \rangle .
\end{eqnarray*}
To characterize the localized orbital state $\left\vert \{A\}\right\rangle $
for each dot, we expand (\ref{V_Exp_3}) to quadratic order and solve the Schr\"{o}dinger equation as though the other
potential wells were absent: 
\begin{equation}
\phi _{A}(\mathbf{r})\equiv \langle \mathbf{r}|A\rangle \equiv \left( \frac{m\omega _{o}}{\pi \hbar }\right) ^{3/4}\exp \left( -\frac{m\omega _{o}}{2\hbar }|\mathbf{r}-\mathbf{A}|^{2}\right).   \label{orb}
\end{equation}
Unless $\alpha $ is small compared to $l^{-2}$, of course, this is
a much coarser approximation than it would be for purely quadratic minima, so we
refine it by centering $\phi _{A}(\mathbf{r})$ and $\phi _{C}(\mathbf{r})$ at the points which
minimize $\langle A|h|A\rangle $ and $\langle C|h|C\rangle $. Because these orbitals overlap at least
slightly for any finite $\omega _{o}$, the states (\ref{psicomp_3}) are not
orthogonal.

We now define $H_{\mathrm{spin}}$ to be the matrix representation of $H$ in
the basis (\ref{psicomp_3}), and expand it in terms of tensor products of
Pauli matrices: 
\begin{equation*}
H_{\mathrm{spin}} = \sum_{i,j,k} c_{ijk} \sigma_{i} \otimes \sigma_{j} \otimes \sigma_{k}.
\end{equation*}
This expansion is always possible, since the set of $n$-fold tensor products
of Pauli matrices constitutes a complete orthonormal basis for the linear
vector space of all $2^{n}\times 2^{n}$ matrices. Because we have written
the basis (\ref{psicomp_3}) in the form $\left\vert s_{A}\right\rangle
\left\vert s_{B}\right\rangle \left\vert s_{C}\right\rangle $, these Pauli
matrices can be associated with spin operators on each of the three quantum
dots. For example, we can write $\sigma _{1}\otimes \sigma _{3}\otimes
\sigma _{0}=2S_{A,x}\otimes 2S_{B,z}\otimes I\equiv 4S_{A,x}S_{B,z}$, where
the notation $S_{W,i}$ means the Pauli operator $\sigma _{i}$ applied to the
electron in the quantum dot at $W$, and where $I$ is the $2\times 2$
identity matrix.  (We exclude $\hbar$ from the definition of the matrices $\sigma_{i}$; thus, the $c_{ijk}$ have the dimensions of energy.)  In the case of an arbitrary $8\times 8$ matrix, 64 complex
numbers would be required to specify our $c_{ijk}$, but the operator (\ref{Hcoord_3}) clearly has certain properties which constrain the values of the
coefficients, such as Hermiticity, reflection symmetry, rotation symmetry,
inversion symmetry, and invariance under permutation of the electrons'
labels. Once these symmetries have been accounted for, the $c_{ijk}$ may be
characterized by just three real quantities: 
\begin{equation}
H_{\mathrm{spin}}=K_{0}+K_{2}[AB](\mathbf{S}_{A}\cdot \mathbf{S}_{B}+\mathbf{%
S}_{B}\cdot \mathbf{S}_{C})+K_{2}[AC]\mathbf{S}_{A}\cdot \mathbf{S}_{C},
\label{HspinJ_3}
\end{equation}
where $\mathbf{S}_{W}\cdot \mathbf{S}_{V}=S_{W,x}S_{V,x}+S_{W,y}S_{V,y}+S_{W,z}S_{V,z}$, and $K_{2}[ij]$ is the
pairwise coupling coefficient between the spins of the electrons in dots $i$
and $j$. Here and elsewhere, we use symmetry considerations to reduce the
number of coupling coefficients in our equations; in this case, the
reflection symmetry of (\ref{V_Exp_3}) through the $x$-$z$
plane implies that $K_{2}[AB]=K_{2}[BC]$. Physically, the constant $K_{2}[AB]
$ quantifies the coupling between adjacent spins, while $K_{2}[AC]$
describes the coupling between the spins at opposite ends of the row.

Defining $\mathbf{S}_{\mathrm{T}} = \mathbf{S}_{A} + \mathbf{S}_{B} + 
\mathbf{S}_{C} $, one finds that 
\begin{equation}  \label{HspinL_3}
H_{\mathrm{spin}} = L_{0} + L_{1}\mathbf{S}_{\mathrm{T}}^{2} +
L_{1}^{\prime}(\mathbf{S}_{A} + \mathbf{S}_{C})^{2},
\end{equation}
where 
\begin{eqnarray}  \label{defn_KJJp}
K_0 &=& L_{0} + \frac{9}{4}L_{1} + \frac{3}{2}L_{1}^{\prime} \\
K_2[AB] &=& 2L_{1}  \notag \\
K_2[AC] &=& 2L_{1} + 2L_{1}^{\prime} .  \notag
\end{eqnarray}

The expansion (\ref{HspinL_3}) reveals that any simultaneous eigenstate of $(\mathbf{S}_{A}+\mathbf{S}_{C})^{2}$ and $\mathbf{S}_{\mathrm{T}}^{2}$ is
also an eigenstate of $H_{\mathrm{spin}}$. We can construct such
simultaneous eigenstates by using the Clebsch-Gordan table twice, first to
combine the spin of the electron in dot $A$ with the spin of the electron in
dot $C$, and then to combine that spin-1 (or spin-0) system with the spin of
the electron in dot $B$: 
\begin{eqnarray}
\left\vert \scriptstyle\frac{3}{2}\displaystyle\;\scriptstyle\frac{3}{2}\displaystyle\;;\;1\right\rangle  &=&|\Psi (\uparrow \uparrow \uparrow
)\rangle   \label{eigenstates_3} \\
\left\vert \scriptstyle\frac{3}{2}\displaystyle\;\scriptstyle\frac{1}{2}\displaystyle\;;\;1\right\rangle  &=&|\Psi (\uparrow \uparrow \downarrow
)\rangle +|\Psi (\uparrow \downarrow \uparrow )\rangle +|\Psi (\downarrow
\uparrow \uparrow )\rangle   \notag \\
\left\vert \scriptstyle\frac{3}{2}\displaystyle\;\scriptstyle\mbox{-}\frac{1}{2}\displaystyle\;;\;1\right\rangle  &=&|\Psi (\downarrow \downarrow
\uparrow )\rangle +|\Psi (\downarrow \uparrow \downarrow )\rangle +|\Psi
(\uparrow \downarrow \downarrow )\rangle   \notag \\
\left\vert \scriptstyle\frac{3}{2}\displaystyle\;\scriptstyle\mbox{-}\frac{3}{2}\displaystyle\;;\;1\right\rangle  &=&|\Psi (\downarrow \downarrow
\downarrow )\rangle   \notag \\
\left\vert \scriptstyle\frac{1}{2}\displaystyle\;\scriptstyle\frac{1}{2}\displaystyle\;;\;1\right\rangle  &=&2|\Psi (\uparrow \downarrow \uparrow
)\rangle -|\Psi (\uparrow \uparrow \downarrow )\rangle -|\Psi (\downarrow
\uparrow \uparrow )\rangle   \notag \\
\left\vert \scriptstyle\frac{1}{2}\displaystyle\;\scriptstyle\mbox{-}\frac{1}{2}\displaystyle\;;\;1\right\rangle  &=&2|\Psi (\downarrow \uparrow
\downarrow )\rangle -|\Psi (\downarrow \downarrow \uparrow )\rangle -|\Psi
(\uparrow \downarrow \downarrow )\rangle   \notag \\
\left\vert \scriptstyle\frac{1}{2}\displaystyle\;\scriptstyle\frac{1}{2}\displaystyle\;;\;0\right\rangle  &=&|\Psi (\uparrow \uparrow \downarrow
)\rangle -|\Psi (\downarrow \uparrow \uparrow )\rangle   \notag \\
\left\vert \scriptstyle\frac{1}{2}\displaystyle\;\scriptstyle\mbox{-}\frac{1}{2}\displaystyle\;;\;0\right\rangle  &=&|\Psi (\downarrow \downarrow
\uparrow )\rangle -|\Psi (\uparrow \downarrow \downarrow )\rangle ,  \notag
\end{eqnarray}
where the indices on the left-hand side denote the values of $S_{\mathrm{T}}$, $S_{\mathrm{T},z}$, and $|\mathbf{S}_{A}+\mathbf{S}_{C}|$ respectively.
Although the states $\left. \mid \Psi (s_{A},s_{B},s_{C})\right\rangle $ are
not orthonormal, the eight states (\ref{eigenstates_3}) are orthogonal, and
they are also eigenvectors of the $8\times 8$ matrix (\ref{HspinL_3}), which
means that $H_{\mathrm{spin}}$ has been diagonalized. To obtain the
parameters $\{L_{0},L_{1},L_{1}^{\prime }\}$, we will choose three
eigenstates with different good quantum numbers, and observe that their
energies can be evaluated either by matrix algebra or by integrating
microscopically over the axes $\mathbf{r}_{i}$ and the spins to compute the
expectation value of (\ref{Hcoord_3}): 
\begin{equation}
\langle \Psi |H_{\mathrm{spin}}|\Psi \rangle =\langle \Psi |H|\Psi \rangle .
\label{E_n}
\end{equation}
Inserting (\ref{HspinL_3}) into the left-hand side, for three distinct
combinations of the good quantum numbers $\{(\mathbf{S}_{A}+\mathbf{S}_{C})^{2},S_{\mathrm{T}}^{2}\}$, yields 
\begin{eqnarray}
\frac{\left\langle \scriptstyle\frac{3}{2}\displaystyle\;\scriptstyle\frac{3}{2}\displaystyle\;;\;1\mid H_{\mathrm{spin}}\mid \scriptstyle\frac{3}{2}\displaystyle\;\scriptstyle\frac{3}{2}\displaystyle\;;\;1\right\rangle }{\left\langle \scriptstyle\frac{3}{2}\displaystyle\;\scriptstyle\frac{3}{2}\displaystyle\;;\;1\mid \scriptstyle\frac{3}{2}\displaystyle\;\scriptstyle\frac{3}{2}\displaystyle\;;\;1\right\rangle } &=&L_{0}+\frac{15}{4}L_{1}+2L_{1}^{\prime }  \label{energyenergyLHS_3} \\
\frac{\left\langle \scriptstyle\frac{1}{2}\displaystyle\;\scriptstyle\frac{1}{2}\displaystyle\;;\;1\mid H_{\mathrm{spin}}\mid \scriptstyle\frac{1}{2}\displaystyle\;\scriptstyle\frac{1}{2}\displaystyle\;;\;1\right\rangle }{\left\langle \scriptstyle\frac{1}{2}\displaystyle\;\scriptstyle\frac{1}{2}\displaystyle\;;\;1\mid \scriptstyle\frac{1}{2}\displaystyle\;\scriptstyle\frac{1}{2}\displaystyle\;;\;1\right\rangle } &=&L_{0}+\frac{3}{4}L_{1}+2L_{1}^{\prime }  \notag \\
\frac{\left\langle \scriptstyle\frac{1}{2}\displaystyle\;\scriptstyle\frac{1}{2}\displaystyle\;;\;0\mid H_{\mathrm{spin}}\mid \scriptstyle\frac{1}{2}\displaystyle\;\scriptstyle\frac{1}{2}\displaystyle\;;\;0\right\rangle }{\left\langle \scriptstyle\frac{1}{2}\displaystyle\;\scriptstyle\frac{1}{2}\displaystyle\;;\;0\mid \scriptstyle\frac{1}{2}\displaystyle\;\scriptstyle\frac{1}{2}\displaystyle\;;\;0\right\rangle } &=&L_{0}+\frac{3}{4}L_{1}, 
\notag
\end{eqnarray}
while the corresponding wave functions (\ref{eigenstates_3}) turn the
right-hand side into 
\begin{widetext}
\begin{eqnarray}
\label{energyenergyRHS_3}
E_{\frac{3}{2},\frac{3}{2};1} &=& \frac{\langle\Psi(\uparrow\uparrow\uparrow) | H | \Psi(\uparrow\uparrow\uparrow)\rangle}{\langle\Psi(\uparrow\uparrow\uparrow) | \Psi(\uparrow\uparrow\uparrow)\rangle} \\
E_{\frac{1}{2},\frac{1}{2};1} &=& \frac{\langle\Psi(\uparrow\uparrow\downarrow) | H | \Psi(\uparrow\uparrow\downarrow)\rangle + 2\langle\Psi(\uparrow\downarrow\uparrow) | H | \Psi(\uparrow\downarrow\uparrow)\rangle - 4\langle\Psi(\uparrow\uparrow\downarrow) | H | \Psi(\uparrow\downarrow\uparrow)\rangle + \langle\Psi(\uparrow\uparrow\downarrow) | H | \Psi(\downarrow\uparrow\uparrow)\rangle}{\langle\Psi(\uparrow\uparrow\downarrow) | \Psi(\uparrow\uparrow\downarrow)\rangle + 2\langle\Psi(\uparrow\downarrow\uparrow) | \Psi(\uparrow\downarrow\uparrow)\rangle - 4\langle\Psi(\uparrow\uparrow\downarrow) | \Psi(\uparrow\downarrow\uparrow)\rangle + \langle\Psi(\uparrow\uparrow\downarrow) | \Psi(\downarrow\uparrow\uparrow)\rangle} \notag \\
E_{\frac{1}{2},\frac{1}{2};0} &=& \frac{\langle\Psi(\uparrow\uparrow\downarrow) | H | \Psi(\uparrow\uparrow\downarrow)\rangle - \langle\Psi(\uparrow\uparrow\downarrow) | H | \Psi(\downarrow\uparrow\uparrow)\rangle}{\langle\Psi(\uparrow\uparrow\downarrow) | \Psi(\uparrow\uparrow\downarrow)\rangle - \langle\Psi(\uparrow\uparrow\downarrow) | \Psi(\downarrow\uparrow\uparrow)\rangle}. \notag
\end{eqnarray}
\end{widetext}
The evaluation of these matrix elements and overlap integrals is a tedious, but straightforward procedure given the microscopic forms of $H$ and $\psi(\mathbf{r})$ in (\ref{Hcoord_3}), (\ref{V_Exp_3}), and (\ref{orb}).  Combining Eqs. (\ref{defn_KJJp}), (\ref{energyenergyLHS_3}), and (\ref{energyenergyRHS_3}), we thus compute $K_{0}$, $K_{2}[AB]$, and $K_{2}[AC]$ in terms of $\omega_o$ and the dimensionless system parameters
\begin{eqnarray}
x_b &\equiv& \frac{\scriptstyle \frac{1}{2} \displaystyle m \omega_o^2 l^2}{\scriptstyle \frac{1}{2} \displaystyle \hbar \omega_o} = \frac{m \omega_o l^2}{\hbar} \label{x_b} \\
x_c &\equiv& \frac{e^2}{\kappa l \hbar \omega_o} \label{x_c} \\
x_v &\equiv& \frac{2 V_0}{\hbar \omega_o}. \label{x_v}
\end{eqnarray}
Physically, the quantity $x_b$ is the ratio of the height of the potential barrier between wells to the energy of the orbital ground state (\ref{orb}), while $x_c$ is the ratio of the equilibrium Coulomb repulsion potential to the energy of the orbital ground state, and $x_v$ is the ratio of the individual well depth $V_0$ to the ground state energy.

Here and in the following section, we have estimated experimentally relevant
values of $x_{b}$ and $x_{c}$ as is done in Ref. \cite{LossDiVincenzo98}. We
assume that the width of the function (\ref{orb}), which is $2\sqrt{\hbar
/m\omega _{o}}$, must be roughly equal to the separation between adjacent
dots $2 l$; using (\ref{x_b}), we conclude that $x_{b}\approx $ 1. For
GaAs heterostructure single dots, $\kappa \approx $ 13, $m^{\ast} \approx $ 0.067 $m_{\mathrm{e}}$, and $\hbar \omega_{o}\approx $ 3 meV, which according to (\ref{x_c}) means that $x_{c}\approx 
$ 1.5. 

A potential of the form (\ref{V_Exp_3}) is most suitable for quantum
computation when $\alpha l^{2}$ is close to 1; if the inverted Gaussian
decays too quickly in space, the spin coupling in the system becomes
negligible, and if it decays too slowly, the local minima in $V$ tend to
coalesce at the center. Using $\scriptstyle\frac{1}{2}\displaystyle\hbar
\omega _{o}\sim $ 1 meV, $V_{0}\approx $ 3 meV \cite{LossDiVincenzo98}, and
our prior estimate of $x_{c}\approx $ 1.5, we obtain the relation $x_{b}\approx x_{v}\sim 3$, by applying (\ref{x_b}), (\ref{x_c}), and (\ref{x_v}). Noting that the parameter $x_{c}$ has very
little influence on any of the coupling constants over physically realistic ranges of $x_{b}$ and $x_{v}$ (and in any
event depends on quantities, such as $\kappa $, which are difficult to tune
experimentally), we henceforth set $x_{c}=1.5$.

\begin{figure}[p]
{\normalsize \includegraphics[scale=0.5]{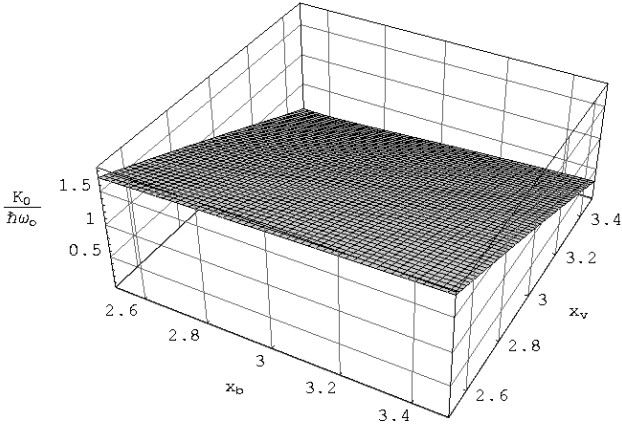} }
\caption{(Color online) Plot of $K_{0}$, the overall energy shift, as a function of
dimensionless barrier height $x_{b}$ and overall well depth $x_{v}$ in the
case of three mutually interacting electrons in a linear geometry. In this
and succeeding figures, the Coulomb repulsion parameter $x_{c}$ is set to
1.5 as in Ref. \protect\cite{MizelLidar}.}
\label{LineExpK}
\end{figure}

\begin{figure}[p]
{\normalsize \includegraphics[scale=0.5]{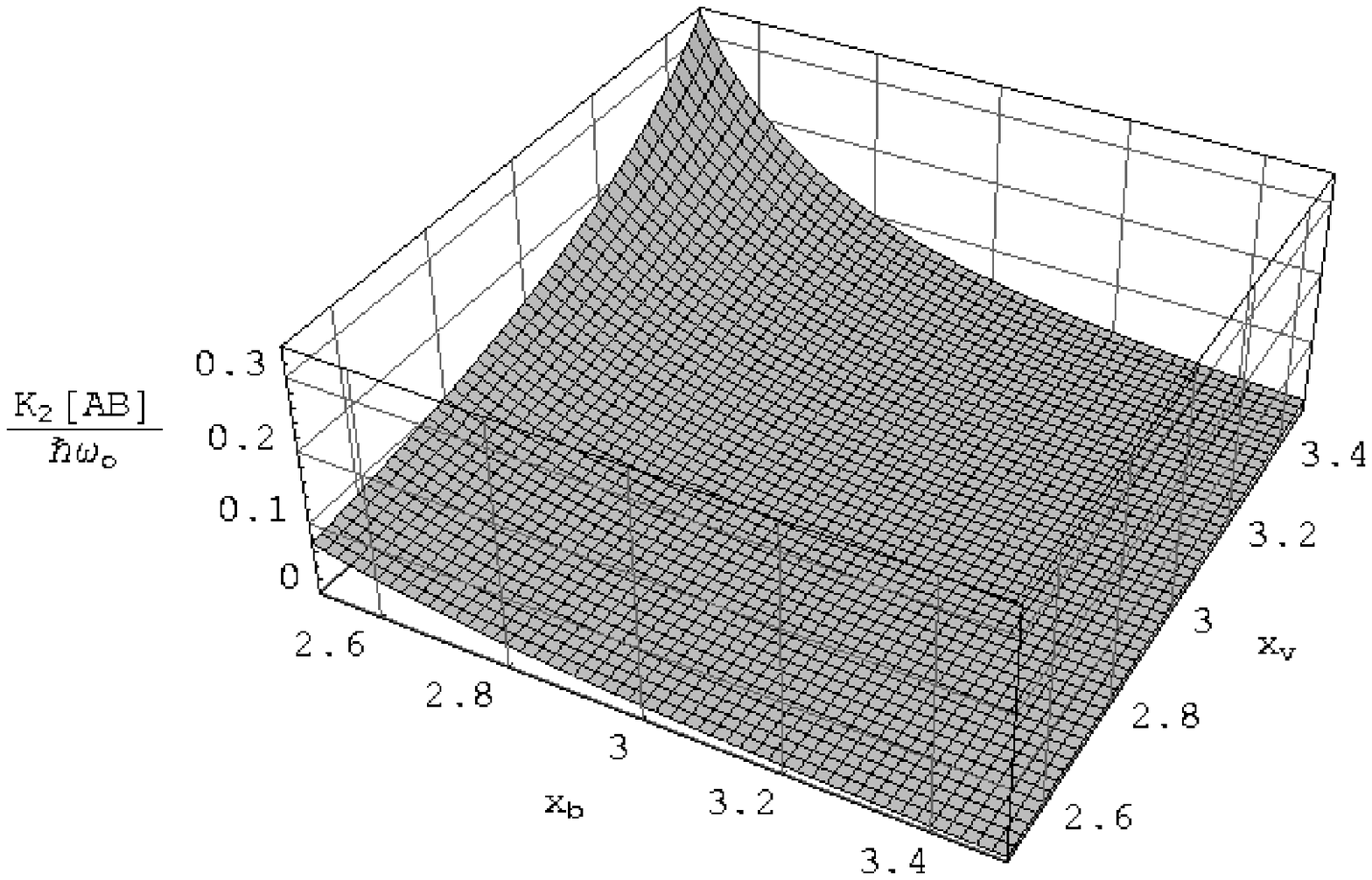} }
\caption{(Color online) Plot of $K_{2}[AB]$, the two-body coupling coefficient for adjacent
dots, as a function of dimensionless barrier height $x_{b}$ and overall well
depth $x_{v}$ in the case of three mutually interacting electrons in a
linear geometry.}
\label{LineExpJ}
\end{figure}

\begin{figure}[p]
{\normalsize \includegraphics[scale=0.5]{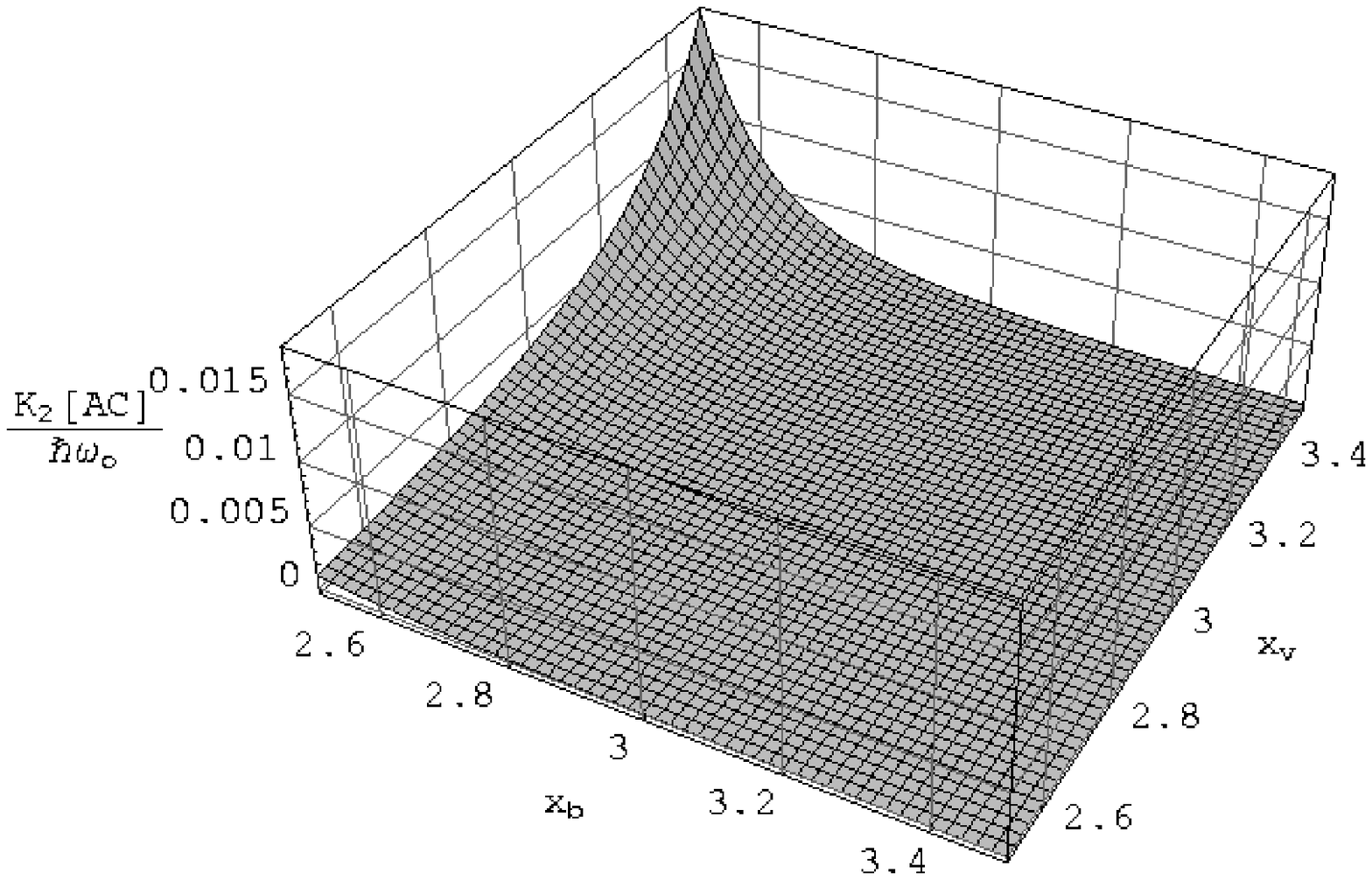} }
\caption{(Color online) Plot of $K_{2}[AC]$, the two-body coupling coefficient for
non-adjacent dots, as a function of dimensionless barrier height $x_{b}$ and
overall well depth $x_{v}$ in the case of three mutually interacting
electrons in a linear geometry.}
\label{LineExpJp}
\end{figure}

Fig.~\ref{LineExpK} shows the energy shift $K_{0}$ as a function of the
system parameters $\{x_{b},x_{v}\}$. As one might expect, this
spin-independent quantity increases with increasing $x_{v}$ and decreasing $x_{b}$ (whenever $\omega _{o}$ decreases, there is greater orbital overlap
and thus more Coulomb repulsion, irrespective of spin state). The coupling
constants $K_{2}[AB]$ and $K_{2}[AC]$ are plotted in Fig.~\ref{LineExpJ}
and Fig.~\ref{LineExpJp} respectively. We notice that they differ (which
rules out the simple Heisenberg form $H_{\mathrm{spin}}=J\sum_{i<j}(\mathbf{S}_{i}\cdot \mathbf{S}_{j})$), and that $K_{2}[AC]$ is only about an order of magnitude
smaller than $K_{2}[AB]$, as we have confirmed by studying $K_{2}[AB](x_{b},x_{v})$ and $K_{2}[AC](x_{b},x_{v})$ on a logarithmic scale. In the context of quantum computation, this
demonstrates that a nearest-neighbor approximation for the coupling between
dots is insufficient (see also Ref. \cite{Scarola:05}, where a
similar conclusion was reported using a low-energy Hubbard model with one electron per site).

\section{Four-Electron Case}

For the case of four quantum dots arranged in a square of side $2l$, our
formalism is more complex in detail but identical in structure. We therefore describe the computation only in outline.

The confining potential in the coordinate Hamiltonian 
\begin{equation}
H=\sum_{i=1}^{4}\left[ \frac{\mathbf{p}_{i}^{2}}{2m}+V(\mathbf{r}_{i})\right]
+\sum_{i<j}\frac{e^{2}}{\kappa \mid \mathbf{r}_{i}-\mathbf{r}_{j}\mid }
\label{Hcoord_4}
\end{equation}
now becomes 
\begin{eqnarray*}
V(\mathbf{r})=-V_{0} &[&e^{-\alpha \mid \mathbf{r}-\mathbf{A}\mid
^{2}}+e^{-\alpha \mid \mathbf{r}-\mathbf{B}\mid ^{2}} \\
&+&e^{-\alpha \mid \mathbf{r}-\mathbf{C}\mid ^{2}}+e^{-\alpha \mid \mathbf{r}-\mathbf{D}\mid ^{2}}],
\end{eqnarray*}
where $\mathbf{A} = (0,2 l,0)$, $\mathbf{B} = (2 l,2 l,0)$, $\mathbf{C} = (2 l,0,0)$, and $\mathbf{D} = (0,0,0)$. Our computational basis consists of 16
fully antisymmetrized vectors of the form 
\begin{eqnarray}
\left\vert \Psi (s_{A},s_{B},s_{C},s_{D})\right\rangle  &=&\sum_{P}\delta
_{P}P[\left\vert A\right\rangle \left\vert B\right\rangle \left\vert
C\right\rangle \left\vert D\right\rangle   \notag  \label{compbasis_4} \\
&&\otimes \left\vert s_{A}\right\rangle \left\vert s_{B}\right\rangle
\left\vert s_{C}\right\rangle \left\vert s_{D}\right\rangle ].
\end{eqnarray}
The form of $\phi (\mathbf{r})$ remains the same; to maintain the required geometrical symmetries, we now shift all four localized orbital wave functions an equal distance toward the point $(l,l,0)$.

Expanding $H$ in terms of products of Pauli matrices, 
\begin{equation*}
H_{\mathrm{spin}} = \sum_{i,j,k,\ell} c_{ijk\ell} \sigma_{i} \otimes \sigma_{j} \otimes \sigma_{k} \otimes \sigma_{l},
\end{equation*}
we discover by applying the symmetries of (\ref{Hcoord_4}) that four-body
terms now appear with nonzero coupling coefficients: 
\begin{eqnarray}
H_{\mathrm{spin}}=K_{0} &+&K_{2}[AB](\mathbf{S}_{A}\cdot \mathbf{S}_{B}+\mathbf{S}_{B}\cdot \mathbf{S}_{C} \notag \\
&&\;\;\;\;\;+\mathbf{S}_{C}\cdot \mathbf{S}_{D}+\mathbf{S}_{D}\cdot \mathbf{S}_{A}) \notag \\
&+&K_{2}[AC](\mathbf{S}_{A}\cdot \mathbf{S}_{C}+\mathbf{S}_{B}\cdot \mathbf{S}_{D}) \notag \\
&+&K_{4}[ABCD][(\mathbf{S}_{A}\cdot \mathbf{S}_{B})(\mathbf{S}_{C}\cdot 
\mathbf{S}_{D}) \notag \\
&&\;\;\;\;\;+(\mathbf{S}_{B}\cdot \mathbf{S}_{C})(\mathbf{S}_{D}\cdot 
\mathbf{S}_{A})] \notag \\
&+&K_{4}[ACBD](\mathbf{S}_{A}\cdot \mathbf{S}_{C})(\mathbf{S}_{B}\cdot 
\mathbf{S}_{D}), \label{HspinJ_4}
\end{eqnarray}
where $K_{4}[ijk\ell]$ is the four-body coupling coefficient among the spins of
the electrons in dots $i$, $j$, $k$, and $\ell$. Physically, the constant $K_{2}[AB]$ describes the pairwise coupling between adjacent spins, while $K_{2}[AC]$ describes the pairwise coupling between non-adjacent spins, $K_{4}[ABCD]$ describes four-body interactions concentrating on pairs of
adjacent spins, and $K_{4}[ACBD]$ describes four-body interactions
concentrating on pairs of non-adjacent spins. We define $S_{\mathrm{T}}=\mathbf{S}_{A}+\mathbf{S}_{B}+\mathbf{S}_{C}+\mathbf{S}_{D}$, which leads us to
\begin{eqnarray}
H_{\mathrm{spin}} &=&L_{0}+L_{1}\mathbf{S}_{\mathrm{T}}^{2}+L_{1}^{\prime }[(\mathbf{S}_{A}+\mathbf{S}_{C})^{2}+(\mathbf{S}_{B}+\mathbf{S}_{D})^{2}] 
\notag  \label{HspinL_4} \\
&&+L_{2}(\mathbf{S}_{\mathrm{T}}^{2})^{2}+L_{2}^{\prime }(\mathbf{S}_{A}+\mathbf{S}_{C})^{2}(\mathbf{S}_{B}+\mathbf{S}_{D})^{2},
\end{eqnarray}
where 
\begin{eqnarray}
K_{0} &=&L_{0}+3L_{1}+3L_{1}^{\prime }+\frac{45}{2}L_{2}+\frac{9}{4}
L_{2}^{\prime }  \label{KJJp_4} \\
K_{2}[AB] &=&2L_{1}+24L_{2}  \notag \\
K_{2}[AC] &=&2L_{1}+2L_{1}^{\prime }+24L_{2}+3L_{2}^{\prime }  \notag \\
K_{4}[ABCD] &=&8L_{2}  \notag \\
K_{4}[ACBD] &=&8L_{2}+4L_{2}^{\prime }.  \notag
\end{eqnarray}
Applying the Clebsch-Gordan table three times creates sixteen simultaneous
eigenstates of $(\mathbf{S}_{A}+\mathbf{S}_{C})^{2}$, $(\mathbf{S}_{B}+\mathbf{S}_{D})^{2}$, and $S_{\mathrm{T}}^{2}$. Inserting five of these states with different quantum numbers into (\ref{E_n}) yields five equations for the five unknowns $\{L_{0},L_{1},L_{1}^{\prime },L_{2},L_{2}^{\prime }\}$ in terms of the
eigenstate energies. As before, these energies may be
expressed in closed form as functions of $x_{b}$, $x_{c}$, and $x_{v}$ by integrating
the right-hand side of (\ref{E_n}) explicitly. 

\begin{figure}[p]
{\normalsize \includegraphics[scale=0.5]{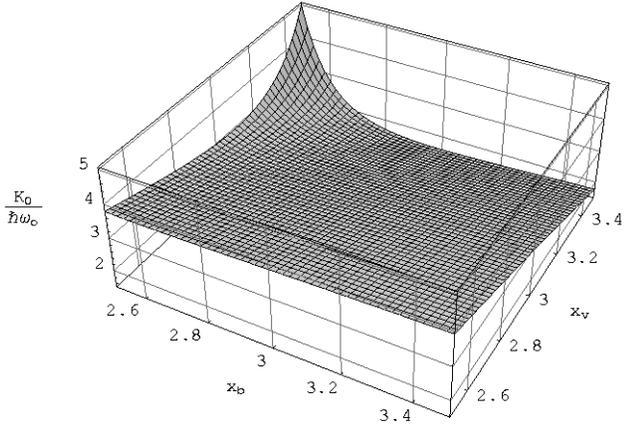} }
\caption{(Color online) Plot of $K_{0}$, the overall energy shift, as a function of
dimensionless barrier height $x_{b}$ and overall well depth $x_{v}$ in the
case of four mutually interacting electrons in a square geometry.}
\label{SquareExpK}
\end{figure}

\begin{figure}[p]
{\normalsize \includegraphics[scale=0.5]{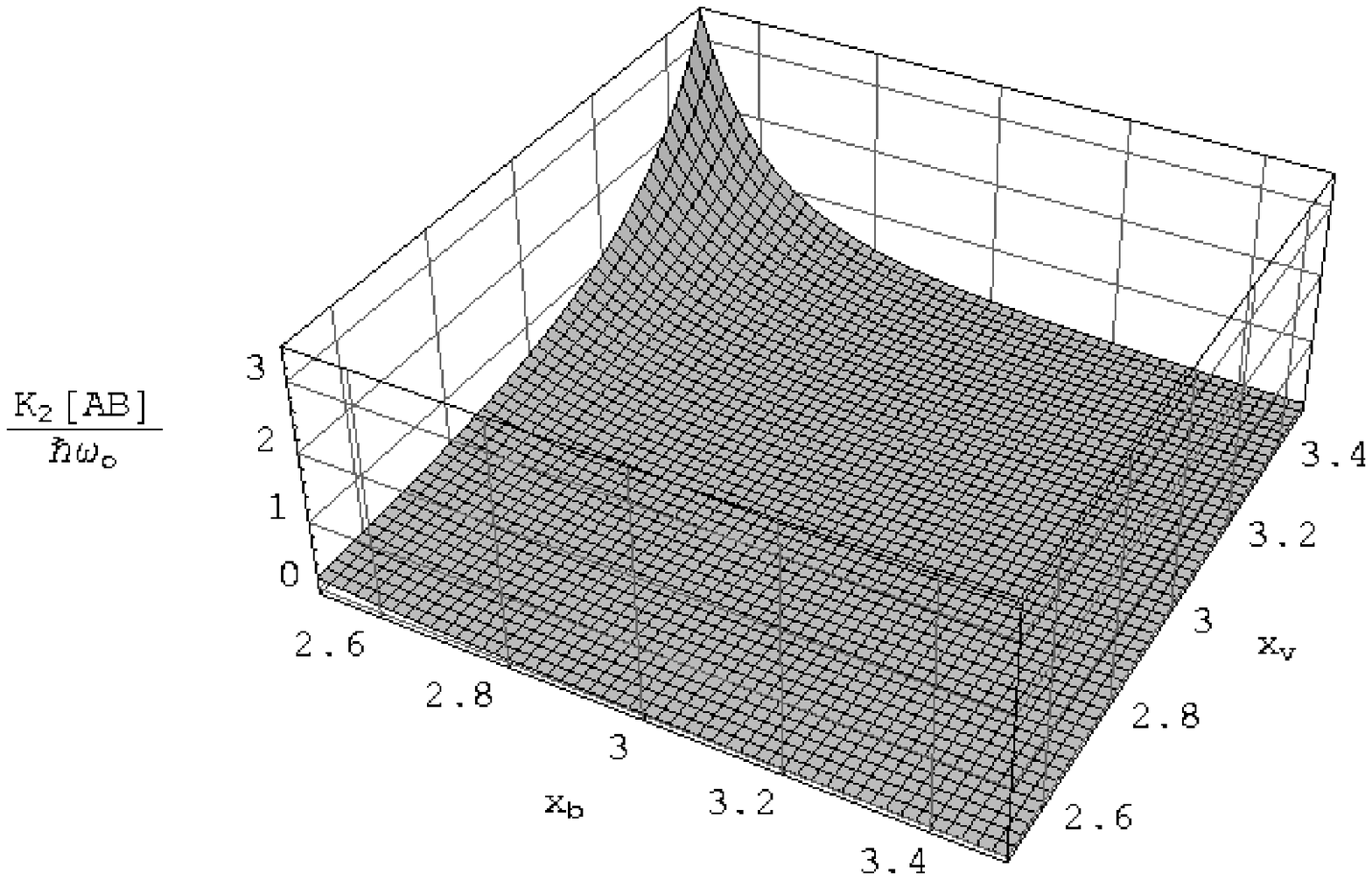} }
\caption{(Color online) Plot of $K_{2}[AB]$, the two-body coupling coefficient for adjacent
dots, as a function of dimensionless barrier height $x_{b}$ and overall well
depth $x_{v}$ in the case of four mutually interacting electrons in a square
geometry.}
\label{SquareExpJ2}
\end{figure}

\begin{figure}[p]
{\normalsize \includegraphics[scale=0.5]{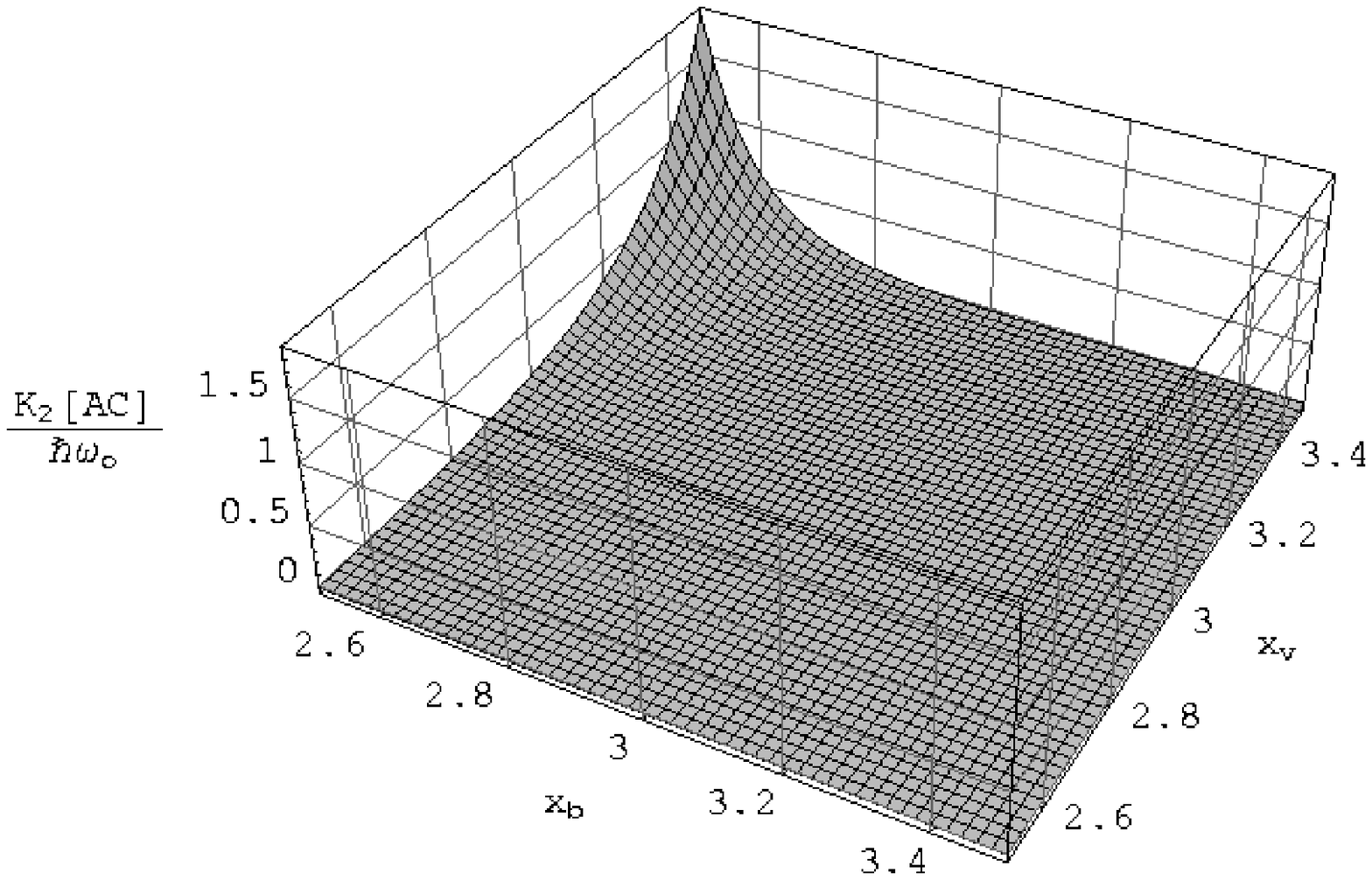} }
\caption{(Color online) Plot of $K_{2}[AC]$, the two-body coupling coefficient for
non-adjacent dots, as a function of dimensionless barrier height $x_{b}$ and
overall well depth $x_{v}$ in the case of four mutually interacting
electrons in a square geometry.}
\label{SquareExpJ2p}
\end{figure}

\begin{figure}[p]
{\normalsize \includegraphics[scale=0.5]{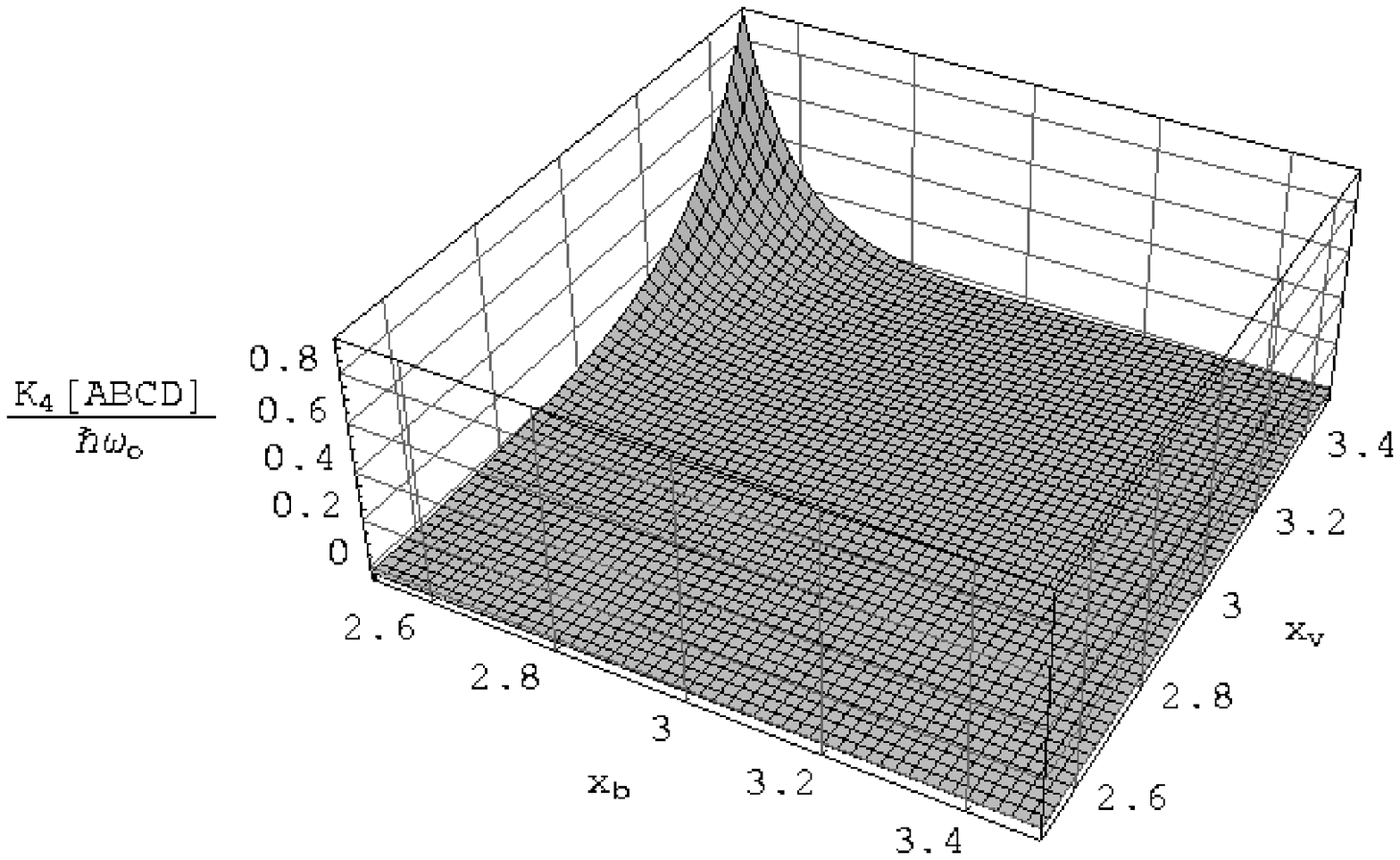} }
\caption{(Color online) Plot of $K_{4}[ABCD]$, the four-body coupling coefficient for pairs
of adjacent dots, as a function of dimensionless barrier height $x_{b}$ and
overall well depth $x_{v}$ in the case of four mutually interacting
electrons in a square geometry.}
\label{SquareExpJ4}
\end{figure}

\begin{figure}[p]
{\normalsize \includegraphics[scale=0.5]{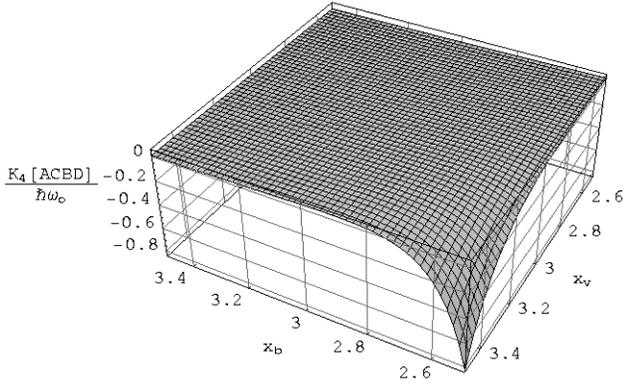} }
\caption{(Color online) Plot of $K_{4}[ACBD]$, the four-body coupling coefficient for pairs
of non-adjacent dots, as a function of dimensionless barrier height $x_{b}$
and overall well depth $x_{v}$ in the case of four mutually interacting
electrons in a square geometry. Note that two of the axis directions are
reversed from the preceding figures.}
\label{SquareExpJ4p}
\end{figure}

The energy shift $K_{0}$ for the square case is plotted in Fig.~\ref{SquareExpK}; as before, this constant is largest for strongly
Coulomb-coupled dots separated by low potential barriers. Figs.~\ref{SquareExpJ2}, ~\ref{SquareExpJ2p}, ~\ref{SquareExpJ4}, and ~\ref{SquareExpJ4p} depict the coupling coefficients $K_{2}[AB]$, $K_{2}[AC]$, $K_{4}[ABCD]$, and $K_{4}[ACBD]$ respectively. The departure from the
pairwise Heisenberg picture is even more pronounced here: we see that for
physically relevant values of the parameters $\{x_{b},x_{v}\}$, the
four-body coefficient $K_{4}[ACBD]$ is of the same order of magnitude as the
two-body coefficient $K_{2}[AC]$, while $K_{4}[ABCD]/K_{2}[AB]\sim $ 0.1, as is
confirmed by plotting $K_{2}[AB](x_{b},x_{v})$, $K_{2}[AC](x_{b},x_{v})$, $K_{4}[ABCD](x_{b},x_{v})$, and $K_{4}[ACBD](x_{b},x_{v})$ on a logarithmic
scale..
Typically, $K_{4}[ACBD]$ is opposite in sign to $K_{2}[AC]$, leading to a
particularly important competition between the two-body and four-body
interactions. 

In order to confirm that the qualitative similarities between our final results and those of Ref. \cite{MizelLidar} were not artifacts of having made two broad changes to $V(\mathbf{r})$ rather than one, we also analyzed both the $N = 3$ and $N = 4$ dot geometries using a confining potential of superposed quadratic minima.  The variation of the coupling coefficients, within experimentally relevant ranges of $x_b$ and $x_c$ (analogous to Figs.~\ref{LineExpK} through~\ref{SquareExpJ4p}), strongly resembled that for the Gaussian potential in all cases.

\section{Computing in the Presence of Four-Body Interactions Using Encoded
Qubits}
\label{encoding}

We have shown that coupling three dots simultaneously quantitatively
modifies the value of the exchange constant, and that coupling four dots
simultaneously switches on a four-body interaction term of the form $K_{4}[ABCD](\mathbf{S}_{A}\cdot \mathbf{S}_{B})(\mathbf{S}_{C}\cdot \mathbf{S}_{D})$ and its permutations. This
conclusion appears to be robust under changes in dot geometry and in the
confining potential. A natural question is whether there exist methods to
cancel the four-body correction. The issue is particularly urgent when one
considers encoded quantum computation (EQC). In many known constructions of
universal gates for EQC \cite{Zanardi:97c,Bacon:99a,DiVincenzo:00a,Kempe:00,Bacon:Sydney,LidarWu:01,Zanardi:99a,WuLidar:01b,WuByrdLidar:02,Lidar:PRA00Exchange,Bacon:01,Lidar:DFSreview,Hsieh:04},
there arises the need to simultaneously couple several spins. One of the
most popular codes, described in detail below, uses four spins per encoded,
or logical qubit \cite{Zanardi:97c,Bacon:99a,Kempe:00,Bacon:Sydney,LidarWu:01,Zanardi:99a,WuLidar:01b,WuByrdLidar:02,Lidar:PRA00Exchange,Bacon:01,Lidar:DFSreview}. For this code, universal computation requires that four spins be coupled
at the same time using pairwise Heisenberg interactions. Hence a
priori it appears that EQC using the four-qubit code suffers from a
fundamental flaw. We now explore whether the four-qubit code may be
implemented in such a way that each four-body coupling is either cancelled
or reduced to an overall phase. Our findings highlight problems that the
four-body terms present in the context of EQC, and also provide an
interesting perspective on how the four-body terms may need to be dealt with
in general.

\subsection{The Code}

Let us descibe the four-spin DFS code, first proposed in Ref. \cite{Zanardi:97c}
in the context of providing immunity against collective decoherence
processes (see Ref. \cite{Lidar:DFSreview} for a review). Let the singlet and
triplet states of two electrons $i,j$ be denoted as 
\begin{eqnarray*}
|s\rangle _{ij} &\equiv &|S=0,m_{S}=0\rangle =\frac{1}{\sqrt{2}}\left( |\Psi
(\uparrow \downarrow )\rangle -|\Psi (\downarrow \uparrow )\rangle \right) 
\\
|t_{-}\rangle _{ij} &\equiv &|S=1,m_{S}=-1\rangle =|\Psi (\downarrow
\downarrow )\rangle  \\
|t_{0}\rangle _{ij} &\equiv &|S=1,m_{S}=0\rangle =\frac{1}{\sqrt{2}}\left(
|\Psi (\uparrow \downarrow )\rangle -|\Psi (\downarrow \uparrow )\rangle
\right)  \\
|t_{+}\rangle _{ij} &\equiv &|S=1,m_{S}=1\rangle =|\Psi (\uparrow \uparrow
)\rangle .
\end{eqnarray*}
Then a single encoded DFS qubit is formed by the two singlets of four spins,
i.e., the two states with zero total spin $S_{\mathrm{T}}=\left| \mathbf{S}_{A}+\mathbf{S}_{B}+\mathbf{S}_{C}+\mathbf{S}_{D} \right|$. These states are formed by combining two
singlets of two pairs of spins ($|0_{L}\rangle $), or triplets of two pairs
of spins, with appropriate Clebsch-Gordan coefficients ($|1_{L}\rangle $): 
\begin{eqnarray}
|0_{L}\rangle  &=&|s\rangle _{AB}\otimes |s\rangle _{CD}  \notag \\
&=&\frac{1}{2}\left( |\Psi (\uparrow \downarrow \uparrow \downarrow )\rangle
+|\Psi (\downarrow \uparrow \downarrow \uparrow )\rangle \right.  \\
&&\;\;\;\left. -|\Psi (\uparrow \downarrow \downarrow \uparrow )\rangle
-|\Psi (\downarrow \uparrow \uparrow \downarrow )\rangle \right) 
\label{eq:0L} \\
|1_{L}\rangle  &=&\frac{1}{\sqrt{3}}\left( |t_{-}\rangle _{AB}\otimes
|t_{+}\rangle _{CD}-|t_{0}\rangle _{AB}\otimes |t_{0}\rangle _{CD}\right.  
\notag \\
&&\;\;\;\left. +|t_{+}\rangle _{AB}\otimes |t_{-}\rangle _{CD}\right)  
\notag \\
&=&\frac{1}{\sqrt{3}}(2|\Psi (\uparrow \uparrow \downarrow \downarrow
)\rangle +2|\Psi (\downarrow \downarrow \uparrow \uparrow )\rangle -|\Psi
(\uparrow \downarrow \downarrow \uparrow )\rangle   \notag \\
&&\;\;\;-|\Psi (\downarrow \uparrow \uparrow \downarrow )\rangle -|\Psi
(\uparrow \downarrow \uparrow \downarrow )\rangle -|\Psi (\downarrow
\uparrow \downarrow \uparrow )\rangle ).  \label{eq:1L}
\end{eqnarray}
As shown in Refs. \cite{Bacon:99a,Kempe:00}, the Heisenberg interaction $\mathbf{S}_{i}\cdot \mathbf{S}_{j}$ can be used all by itself to implement universal quantum
computation on this type of code. The Heisenberg interaction is closely
related to the exchange operator $E_{ij}$, defined as 
\begin{equation}
E_{ij}=\left( 
\begin{array}{cccc}
1 &  &  &  \\ 
& 0 & 1 &  \\ 
& 1 & 0 &  \\ 
&  &  & 1
\end{array}
\right)   \label{eq:Ematrix}
\end{equation}
via $E_{ij}=\frac{1}{2}(4\mathbf{S}_{i}\cdot \mathbf{S}_{j}+I)$. The difference in their
action as gates is only a phase, so that we will use $E_{ij}$ and $\mathbf{S}_{i}\cdot \mathbf{S}_{j}$ interchangeably from now on and write $E_{ij}\simeq
\mathbf{S}_{i}\cdot \mathbf{S}_{j}$. The $E_{ij}$ have a simple action on the electronic spin
up/down states, as seen from the matrix representation (\ref{eq:Ematrix}):\
the states $|00\rangle $ and $|11\rangle $ are invariant, whereas $
|01\rangle $ and $|10\rangle $ are exchanged. Using this, it is simple to
show that, in the $\{|0_{L}\rangle ,|1_{L}\rangle \}$ basis, the exchange
operators can be written as \cite{Lidar:PRA00Exchange,Kempe:00} 
\begin{eqnarray}
E_{AB} &=&E_{CD}=\left( 
\begin{array}{cc}
-1 & 0 \\ 
0 & 1
\end{array}
\right) =-\bar{Z} \\
E_{AC} &=&E_{BD}=\frac{\sqrt{3}}{2}\bar{X}+\frac{1}{2}\bar{Z}  \notag \\
E_{AD} &=&E_{BC}=-\frac{\sqrt{3}}{2}\bar{X}+\frac{1}{2}\bar{Z},  \notag
\label{eq:Pauli-DFS}
\end{eqnarray}
where $\bar{X},\bar{Z}$ are the encoded Pauli matrices $\sigma _{x},\sigma
_{z}$, i.e., the Pauli matrices acting on the $|0_{L}\rangle ,|1_{L}\rangle $
states. It follows from the Euler angle formula, $e^{-i\omega \mathbf{n}\cdot \boldsymbol{\sigma }}=e^{-i\beta \sigma _{z}}e^{-i\theta \sigma
_{x}}e^{-i\alpha \sigma _{z}}$ (a rotation by angle $\omega $ about the axis 
$\mathbf{n}$, given in terms of three successive rotations about the $z$ and 
$x$ axes), that one can perform all single encoded-qubit operations on the
DFS states, simply by switching the exchange interaction on and off. Note that the Euler angle formula is satisfied by any pair of non-parallel axes, but
orthogonal axes may be more convenient. One can obtain an encoded $\sigma
_{x}$ operation by switching on two interactions simultaneously for the
appropriate time intervals: 
\begin{equation*}
\bar{X}=-2\left( E_{AC}+\frac{1}{2}E_{AB}\right) /\sqrt{3}=\left(
E_{AC}-E_{AD}\right) /\sqrt{3}.
\end{equation*}
Use of the Euler angle formula requires a Hamiltonian which is a sum of
exchange terms with controllable coefficients $J_{ij}(t)$: 
\begin{equation*}
H_{\mathrm{S}}=\sum_{i\neq j}J_{ij}(t)E_{ij}.
\end{equation*}
This is achievable, e.g., by using local magnetic fields \cite{LossDiVincenzo98,Burkard:99,Burkard:00,Hu:99,Hu:01b}, by ferroelectric gates \cite{Levy:01}, or by optical rectification \cite{Levy:02}. It is
important to emphasize that the last two methods \cite{Levy:01,Levy:02} do
not require magnetic field control, hence overcome at least in part the
problems with EQC\ raised in Refs. \cite{Scarola:04,Scarola:05}. This is an
important advantage with regards to EQC, which renders these electrical-only
type control methods distinctly preferable to those using magnetic fields.
However, residual magnetic fields, e.g., due to nuclear spin impurities, do
remain a problem, especially in the group III-V semiconductors, such as GaAs 
\cite{sousa:115322}. In silicon-based architectures
this problem can be minimized by isotopic purification \cite{Yablonovitch:03}.

\subsection{Effect of the Four-Body Terms on a Single Encoded Qubit}

Let us now consider how the four-body terms act on the DFS code. Using the
results above, we find that
\begin{equation*}
(\mathbf{S}_{A}\cdot \mathbf{S}_{B})(\mathbf{S}_{C}\cdot \mathbf{S}_{D})\simeq E_{AB}E_{CD}=(-\bar{Z})^{2}=I,
\end{equation*}
where $I$ is the identity operator. Also,
\begin{eqnarray*}
E_{AC}E_{BD} &=&\left( \frac{\sqrt{3}}{2}\bar{X}+\frac{1}{2}\bar{Z}\right)
^{2} \\
&=&\frac{1}{4}\left( 3I+I+\sqrt{3}(\bar{X}\bar{Z}+\bar{Z}\bar{X})\right) =I,
\end{eqnarray*}
and similarly $E_{AD}E_{BC}=I$. Thus all fourth-order terms $(\mathbf{S}_{i}\cdot
\mathbf{S}_{j})(\mathbf{S}_{k}\cdot \mathbf{S}_{l})\propto I$ as long we restrict their action to the
subspace encoding one qubit. This implies that the encoding into the 4-qubit
DFS\ is immune to the fourth-order terms. In other words, when this encoding
is used, the problem of the computational errors induced by the undesired
fourth-order terms simply disappears, as long as we restrict our attention
to a single encoded qubit.

\subsection{Two Encoded Qubits}

We must also be able to couple encoded qubits via a non-trivial gate such as
controlled-phase:$\ CP=\mathrm{diag}(-1,1,1,1)$. This is one way to satisfy the requirements for universal quantum computation \cite{Barenco:95a}, though it is also possible to complete the set of
single-qubit gates by measurements \cite{KLM-Nature}. Two encoded qubits of the form (\ref{eq:0L}), (\ref{eq:1L}) occupy a
four-dimensional subspace of the zero total spin subspace of $8$ spins. The zero total spin
subspace is $14$-dimensional. A very useful graphical way of seeing this, introduced in Ref. \cite{Kempe:00} but also known as a Bratteli diagram, is depicted in Fig.~\ref{fig:DFS}.

As more spins are added (horizontal axis), there are more possibilities for
constructing a state with given total spin (vertical axis). In the case of
four spins there are two paths leading from the origin to $S_{\mathrm{T}}=0$; these correspond exactly to the $|0_{L}\rangle $ and $|1_{L}\rangle $ code
states. For eight spins there are $14$ such paths. Only four of these
correspond to the four basis states $|0_{L}0_{L}\rangle ,|0_{L}1_{L}\rangle
,|1_{L}0_{L}\rangle ,|1_{L}1_{L}\rangle $. It is convenient to label paths
according to the intermediate total spin: the state $|S_{1},S_{2},S_{3},S_{4},S_{5},S_{6},S_{7},S_{8}\rangle $, where $S_{k}$ is
the total spin of $k$ spin-1/2 particles, uniquely corresponds to a path in
Fig.~\ref{fig:DFS} (we omit the origin in this notation), and the $S_{k}$
form a complete set of commuting observables \cite{Kempe:00}. E.g., 
\begin{widetext}
\begin{eqnarray*}
|0_{L}0_{L}\rangle &=&|1/2,0,1/2,0,1/2,0,1/2,0\rangle =
\begin{tabular}{llllllll}
$\rule[-3mm]{0mm}{4mm} \nearrow $ & $\searrow $ & $\nearrow $ & $\searrow $ & $\nearrow $ & $
\searrow $ & $\nearrow $ & $\searrow $
\end{tabular}
\\
|0_{L}1_{L}\rangle &=&|1/2,0,1/2,0,1/2,1,1/2,0\rangle =
\begin{tabular}{cccccccc}
&  &  &  &  & $\nearrow $ & $\searrow $ &  \\ 
$\nearrow $ & $\searrow $ & $\nearrow $ & $\searrow $ & $\nearrow $ &  &  & $
\searrow $
\end{tabular}
\\
|1_{L}0_{L}\rangle &=&|1/2,1,1/2,0,1/2,0,1/2,0\rangle =
\begin{tabular}{cccccccc}
& $\nearrow $ & $\searrow $ &  &  &  &  &  \\ 
$\nearrow $ &  &  & $\searrow $ & $\nearrow $ & $\searrow $ & $\nearrow $ & $
\searrow $
\end{tabular}
\\
|1_{L}1_{L}\rangle &=&|1/2,1,1/2,0,1/2,1,1/2,0\rangle =
\begin{tabular}{cccccccc}
& $\nearrow $ & $\searrow $ &  &  & $\nearrow $ & $\searrow $ &  \\ 
$\nearrow $ &  &  & $\searrow $ & $\nearrow $ &  &  & $\searrow $
\end{tabular}
.
\end{eqnarray*}
\end{widetext}On the right we have indicated the path in Fig.~\ref{fig:DFS}
corresponding to each state. The other $10$ states with zero total spin can
be similarly described. Thus the set of $14$ states $\{|S_{1},S_{2},S_{3},S_{4},S_{5},S_{6},S_{7},0\rangle \}$ forms a basis for
the subspace of zero total spin of $8$ spin-1/2 particles. Henceforth we
will find it convenient to represent exchange operators in this basis. We
will order the $14$ basis states as follows: first the four code states $|0_{L}0_{L}\rangle ,|0_{L}1_{L}\rangle ,|1_{L}0_{L}\rangle
,|1_{L}1_{L}\rangle $ as above, then 
\begin{eqnarray*}
&&|1/2,0,1/2,1,1/2,0,1/2,0\rangle ,|1/2,1,1/2,1,1/2,0,1/2,0\rangle , \\
&&|1/2,0,1/2,1,1/2,1,1/2,0\rangle ,|1/2,1,1/2,1,1/2,1,1/2,0\rangle , \\
&&|1/2,0,1/2,1,3/2,1,1/2,0\rangle ,|1/2,1,3/2,1,1/2,0,1/2,0\rangle , \\
&&|1/2,1,3/2,1,3/2,1,1/2,0\rangle ,|1/2,1,3/2,2,3/2,1,1/2,0\rangle , \\
&&|1/2,1,1/2,1,3/2,1,1/2,0\rangle ,|1/2,1,3/2,1,1/2,1,1/2,0\rangle .
\end{eqnarray*}%
E.g., in this basis the operator $E_{DE}$ has the representation\footnote{{\normalsize All matrix calculations reported here were performed with
Mathematica.}} 
\begin{widetext}
\begin{equation*}
E_{DE}=\left( 
\begin{array}{cccccccccccccc}
\frac{1}{2} &  &  &  & \frac{\sqrt{3}}{2} &  &  &  &  &  &  &  &  &  \\ 
& \frac{1}{2} &  &  &  &  & \frac{\sqrt{3}}{2} &  &  &  &  &  &  &  \\ 
&  & \frac{1}{2} &  &  & \frac{\sqrt{3}}{2} &  &  &  &  &  &  &  &  \\ 
&  &  & \frac{1}{2} &  &  &  & \frac{\sqrt{3}}{2} &  &  &  &  &  &  \\ 
\frac{\sqrt{3}}{2} &  &  &  & -\frac{1}{2} &  &  &  &  &  &  &  &  &  \\ 
&  & \frac{\sqrt{3}}{2} &  &  & -\frac{1}{2} &  &  &  &  &  &  &  &  \\ 
& \frac{\sqrt{3}}{2} &  &  &  &  & -\frac{1}{2} &  &  &  &  &  &  &  \\ 
&  &  & \frac{\sqrt{3}}{2} &  &  &  & -\frac{1}{2} &  &  &  &  &  &  \\ 
&  &  &  &  &  &  &  & 1 &  &  &  &  &  \\ 
&  &  &  &  &  &  &  &  & 1 &  &  &  &  \\ 
&  &  &  &  &  &  &  &  &  & \frac{1}{4} & \frac{\sqrt{15}}{4} &  &  \\ 
&  &  &  &  &  &  &  &  &  & \frac{\sqrt{15}}{4} & -\frac{1}{4} &  &  \\ 
&  &  &  &  &  &  &  &  &  &  &  & 1 &  \\ 
&  &  &  &  &  &  &  &  &  &  &  &  & 1
\end{array}
\right).
\end{equation*}
\end{widetext}Recall that the first four rows refer to the code space. It is
then clear that $E_{DE}$ mixes the code space with four of the remaining ten
states that have zero total spin. This is a general feature of all exchange
operators acting on two code blocks simultaneously. For this reason it is
impossible to couple two code blocks in one step, while preserving the code
space.

\subsection{Enacting an Encoded Controlled-Phase Gate}

For the 4-qubit code above, procedures implementing a $CP$ gate were first
given in Refs. \cite{Bacon:99a,Kempe:00}. Recently, Bacon \cite[App. E]{Bacon:thesis} found a simplified scheme which is a useful starting point for
our purposes. Bacon's implementation of a $CP$ gate between two pairs of
4-qubit blocks (qubits $A$-$D$ and qubits $E$-$H$) involves a sequence of $14 $ elementary gates, each of which requires that several simultaneous
exchange interactions be switched on and off. We will here take the approach
of utilizing Bacon's construction, while making some modifications due to
the appearance of three- and four-body corrections. For ease of
visualization, we will assume that the two blocks are squares of side $ 2l $ and that dots $D$ and $E$ are separated by a distance $ 2l $, although
nearly all of the following calculations are independent of the exact
spatial relationship between the blocks.

The gates are (adapting the notation of \cite[App. E]{Bacon:thesis})

\begin{eqnarray*}
U_{1} &=&\exp \left[ \frac{i\pi }{\sqrt{3}}\left( E_{DE}+\frac{1}{2}\sum_{A=i<j}^{D}E_{ij}\right) \right]  \\
U_{2} &=&\exp \left[ \frac{i\pi }{4\sqrt{2}}\left( -3E_{EF}-\frac{2}{3}\left( E_{FG}+E_{FH}+E_{GH}\right) \right) \right]  \\
U_{3} &=&\exp \left[ \frac{i\pi }{4\sqrt{2}}\left( -3E_{CD}-\frac{2}{3}\left( E_{AB}+E_{AC}+E_{BC}\right) \right) \right]  \\
U_{5} &=&\exp \left[ \frac{i\pi }{\sqrt{3}}\left( E_{FG}+\frac{1}{2}E_{GH}\right) \right]  \\
U_{6} &=&\left( U_{A}U_{B}U_{A}^{\dagger }U_{B}^{\dagger }\right) ^{2},
\end{eqnarray*}
where 
\begin{eqnarray*}
U_{A} &=&\exp \left[ -\frac{i}{2}\cos ^{-1}(-1/3)E_{DE}\right]  \\
U_{B} &=&\exp \left[ -\frac{i\pi }{2}\sum_{A=i<j}^{D}E_{ij}\right] .
\end{eqnarray*}
In terms of these gates, the controlled-phase gate\ can be written as
\begin{equation*}
CP=U_{1}^{\dagger }(U_{2}^{\dagger }U_{3}^{\dagger })U_{5}^{\dagger
}U_{6}U_{5}(U_{3}U_{2})U_{1},
\end{equation*}
where $U_{3}U_{2}$ can be executed in one step since by inspection the two
gates operate on the two blocks separately (and identically). This gate
sequence operates in the entire $14$-dimensional subspace of $S_{\mathrm{T}}=0$ states of $8$ spins:\ the code space is left after application of $U_{1}
$, but is returned to at the end of the sequence, when $U_{1}^{\dagger }$ is
applied. Hence our single-qubit considerations above do not apply:
even if a four-body interaction acts as the identity operator on a single encoded qubit,
it may act non-trivially in the larger $S_{\mathrm{T}}=0$ space. We must
therefore carefully analyze the action of this gate sequence in light of the
three- and four-body corrections.

The key point in Bacon's construction of the gate sequence is to ensure that
each gate acts \textquotedblleft classically,\textquotedblright\ i.e., it
only couples a given $S_{\mathrm{T}}=0$ basis state to another, without
creating superpositions of such basis states (that the gates above act in
this manner is not at all simple to see directly, but is the reason for the
particular choice of angles in the gates). Here we will show that, in order
to still satisfy this key point, it is necessary to tune the four-body
exchange coupling constants. Thus, to enact a $CP$ gate in the presence of
four-body interactions, there needs to be sufficient flexibility in tuning
the four-body coupling. We note that there are other ways to obtain a $CP$ gate \cite{Hsieh:04}; our point here is mostly to explore the implications of the four-body terms in a context of some general interest. Let us now consider each of the gates in detail, in increasing order of complexity.

\subsubsection{The $U_{A}$ Gate}

$U_{A}$ only involves a single exchange interaction, and so is unmodified in the
presence of the three- and four-body corrections:

\begin{widetext}
\begin{eqnarray*}
U_{A}^{\prime } &=&U_{A}=\exp \left[ -\frac{i}{2}\cos ^{-1}(-1/3)E_{DE}
\right] \\
&=&\left( 
\begin{array}{cccccccccccccc}
\alpha &  &  &  & \frac{1}{i\sqrt{2}} &  &  &  &  &  &  &  &  &  \\ 
& \alpha &  &  &  &  & \frac{1}{i\sqrt{2}} &  &  &  &  &  &  &  \\ 
&  & \alpha &  &  & \frac{1}{i\sqrt{2}} &  &  &  &  &  &  &  &  \\ 
&  &  & \alpha &  &  &  & \frac{1}{i\sqrt{2}} &  &  &  &  &  &  \\ 
\frac{1}{i\sqrt{2}} &  &  &  & {{\alpha }^{\ast }} &  &  &  &  &  &  &  &  & 
\\ 
&  & \frac{1}{i\sqrt{2}} &  &  & {{\alpha }^{\ast }} &  &  &  &  &  &  &  & 
\\ 
& \frac{1}{i\sqrt{2}} &  &  &  &  & {{\alpha }^{\ast }} &  &  &  &  &  &  & 
\\ 
&  &  & \frac{1}{i\sqrt{2}} &  &  &  & {{\alpha }^{\ast }} &  &  &  &  &  & 
\\ 
&  &  &  &  &  &  &  & \beta &  &  &  &  &  \\ 
&  &  &  &  &  &  &  &  & \beta &  &  &  &  \\ 
&  &  &  &  &  &  &  &  &  & \frac{1}{\sqrt{3}}+\frac{1}{2i\sqrt{6}} & \frac{
1}{2i}\sqrt{\frac{5}{2}} &  &  \\ 
&  &  &  &  &  &  &  &  &  & \frac{1}{2i}\sqrt{\frac{5}{2}} & \frac{1}{
\sqrt{3}}-\frac{1}{2i\sqrt{6}} &  &  \\ 
&  &  &  &  &  &  &  &  &  &  &  & \beta &  \\ 
&  &  &  &  &  &  &  &  &  &  &  &  & \beta
\end{array}
\right) ,
\end{eqnarray*}
\end{widetext}where $\alpha =\frac{\sqrt{2}-i}{\sqrt{6}}$ and $\beta
=e^{-i/(2\cos ^{-1}(-3))}$.

\subsubsection{The $U_{5}$ Gate}

$U_{5}$ involves dots $F$, $G$, and $H$, with dots $F$ and $H$
simultaneously coupled to $G$, and hence experiences a three-body correction
to the exchange constants. In addition, a coupling between dot $F$ and dot $H $ will arise, which forces our modified $U_{5}$ gate to have the form

\begin{widetext}
\begin{eqnarray*}
U_{5}^{\prime } &=& \exp \left[ \frac{i\pi }{\sqrt{3}}\left( E_{FG}+
\frac{1}{2}E_{GH} + J_5^{\prime} E_{FH}\right) \right] \\
&=&\left( 
\begin{array}{cccccccccccccc}
p    & \Phi &      &      &      &      &      &      &      &      &      &      &      &        \\ 
\Phi & p^{*}&      &      &      &      &      &      &      &      &      &      &      &        \\ 
     &      & p    & \Phi &      &      &      &      &      &      &      &      &      &        \\ 
     &      & \Phi & p^{*}&      &      &      &      &      &      &      &      &      &        \\ 
     &      &      &      & p    &      & \Phi &      &      &      &      &      &      &        \\ 
     &      &      &      &      & p    &      & \Phi &      &      &      &      &      &        \\ 
     &      &      &      & \Phi &      & p^{*}&      &      &      &      &      &      &        \\ 
     &      &      &      &      & \Phi &      & p^{*}&      &      &      &      &      &        \\ 
     &      &      &      &      &      &      &      & e^{\frac{i \pi (2 J_5 + 3)}{2 \sqrt{3}}}     &      &      &      &      &        \\ 
     &      &      &      &      &      &      &      &      & p    &      &      &      & \Phi   \\ 
     &      &      &      &      &      &      &      &      &      & e^{\frac{i \pi (2 J_5 + 3)}{2 \sqrt{3}}}     &      &      &        \\ 
     &      &      &      &      &      &      &      &      &      &      & e^{\frac{i \pi (2 J_5 + 3)}{2 \sqrt{3}}}     &      &        \\ 
     &      &      &      &      &      &      &      &      &      &      &      & e^{\frac{i \pi (2 J_5 + 3)}{2 \sqrt{3}}}     &        \\ 
     &      &      &      &      &      &      &      &      & \Phi &      &      &      & p^{*}     
\end{array}
\right),
\end{eqnarray*}
\end{widetext}where 
\begin{eqnarray*}
\Lambda  &=&\sqrt{\scriptstyle\frac{4}{3}\displaystyle(J_{5}^{\prime
})^{2}-2J_{5}^{\prime }+1} \\
p &=&\cos \left[ \frac{\pi }{2} \Lambda \right] +\frac{iJ_{5}^{\prime }}{\sqrt{3}\Lambda }\sin \left[ \frac{\pi }{2}\Lambda \right]  \\
\Phi  &=&\frac{i}{J_{5}^{\prime }}(1-J_{5}^{\prime })\sin \left[ \frac{\pi }{2}\Lambda \right] 
\end{eqnarray*}
It is seen that this gate operates \textquotedblleft
classically,\textquotedblright\ and is non-diagonal, only when $J_{5}^{\prime }$ is either $0$ or chosen such that $\Lambda $ is an even
integer. In both cases, we recover Bacon's functional form:

\footnotesize 
\begin{equation*}
U_{5}=\left( 
\begin{array}{cccccccccccccc}
& i &  &  &  &  &  &  &  &  &  &  &  &  \\ 
i &  &  &  &  &  &  &  &  &  &  &  &  &  \\ 
&  &  & i &  &  &  &  &  &  &  &  &  &  \\ 
&  & i &  &  &  &  &  &  &  &  &  &  &  \\ 
&  &  &  &  &  & i &  &  &  &  &  &  &  \\ 
&  &  &  &  &  &  & i &  &  &  &  &  &  \\ 
&  &  &  & i &  &  &  &  &  &  &  &  &  \\ 
&  &  &  &  & i &  &  &  &  &  &  &  &  \\ 
&  &  &  &  &  &  &  & e^{i\frac{\sqrt{3}\pi }{2}} &  &  &  &  &  \\ 
&  &  &  &  &  &  &  &  &  &  &  &  & i \\ 
&  &  &  &  &  &  &  &  &  & e^{i\frac{\sqrt{3}\pi }{2}} &  &  &  \\ 
&  &  &  &  &  &  &  &  &  &  & e^{i\frac{\sqrt{3}\pi }{2}} &  &  \\ 
&  &  &  &  &  &  &  &  &  &  &  & e^{i\frac{\sqrt{3}\pi }{2}} &  \\ 
&  &  &  &  &  &  &  &  & i &  &  &  & 
\end{array}
\right) .
\end{equation*}
\normalsize

\subsubsection{The $U_{B}$ Gate}

Bacon's $U_{B}$ gate is 
\begin{eqnarray*}
U_{B} &=&\exp \left[ -\frac{i\pi }{2}\left( \sum_{A=i<j}^{D}E_{ij}\right) \right] \\
&=&\mathrm{diag}(1,1,1,1,-1,-1,-1, \\
&&\;\;\;\;-1,-1,-1,-1,-1,-1,-1).
\end{eqnarray*}
$U_{B}$ involves coupling between four dots, so experiences both
quantitative three-body corrections and a four-body qualitative correction.
Since the four spins in $U_{B}$ are coupled symmetrically, the form of the
four-body correction must also be symmetric: 
\begin{eqnarray*}
U_{B}^{\prime } &=&\exp \left[ -\frac{i\pi }{2}\left(
\sum_{A=i<j}^{D}E_{ij}+J_{B}^{\prime }\left( E_{AB}E_{CD}\right. \right.
\right. \\
&&\left. \left. \left. \;\;\;\;\;+E_{AC}E_{BD}+E_{AD}E_{BC}\right) \right) 
\right] \\
&=&\mathrm{diag}(\gamma ^{-3},\gamma ^{-3},\gamma ^{-3},\gamma ^{-3},-\gamma
,-\gamma ,-\gamma , \\
&&\;\;\;\;-\gamma ,-\gamma ,-\gamma ,-\gamma ,-\gamma ^{-3},-\gamma ,-\gamma
),
\end{eqnarray*}
where $\gamma =e^{\frac{i\pi }{2}{J_{B}^{\prime }}}$. Note that because $U_{B}^{\prime }$ forms part of the $U_{6}^{\prime }$ gate, that acts on $S_{\mathrm{T}}=0$ states outside of the code space, the action of the four-body
terms in it is nontrivial for arbitrary $J_{B}^{\prime }$. However, upon
setting $J_{B}^{\prime }$\ to any integer value we recover $U_{B}$, up to an
overall phase.

\subsubsection{The $U_{2},U_{3}$ Gates}

$U_{2},U_{3}$ similarly involve coupling between four dots inside a fixed
code block, so also experience both quantitative three-body corrections and
a four-body qualitative correction. In this case the Heisenberg couplings
are not symmetric, so we do not assume that the four-body terms are all
turned on with equal coupling constants: 
\footnotesize
\begin{eqnarray*}
U_{2}^{\prime } &=&\exp \left[ \frac{i\pi }{4\sqrt{2}}\left( -3E_{EF}-\frac{2}{3}\left( E_{FG}+E_{FH}+E_{GH}\right) \right. \right. \\
&&\left. \left. \;\;\;\;\;\;\;\;\;+J_{2}^{\prime }E_{EF}E_{GH}+J_{2}^{\prime
\prime }\left( E_{AC}E_{BD}+E_{EH}E_{FG}\right) \right) \right] \\
&=&\left( 
\begin{array}{cccccccccccccc}
\delta &  &  &  &  &  &  &  &  &  &  &  &  &  \\ 
& \epsilon &  &  &  &  &  &  &  &  &  &  &  &  \\ 
&  & \delta &  &  &  &  &  &  &  &  &  &  &  \\ 
&  &  & \epsilon &  &  &  &  &  &  &  &  &  &  \\ 
&  &  &  & \zeta &  &  &  &  &  &  &  &  &  \\ 
&  &  &  &  & \zeta &  &  &  &  &  &  &  &  \\ 
&  &  &  &  &  & \eta &  & \rho &  &  &  &  &  \\ 
&  &  &  &  &  &  & \eta &  &  &  &  & \rho &  \\ 
&  &  &  &  &  & \rho &  & \eta ^{*} &  &  &  &  &  \\ 
&  &  &  &  &  &  &  &  & \zeta &  &  &  &  \\ 
&  &  &  &  &  &  &  &  &  & \eta ^{*} &  &  & \rho \\ 
&  &  &  &  &  &  &  &  &  &  & e^{i\pi \frac{\left( -5+J_{2}^{\prime
}+2J_{2}^{\prime \prime }\right) }{4\sqrt{2}}} &  &  \\ 
&  &  &  &  &  &  & \rho &  &  &  &  & \eta ^{*} &  \\ 
&  &  &  &  &  &  &  &  &  & \rho &  &  & \eta
\end{array}
\right) ,
\end{eqnarray*}
\normalsize
and 
\footnotesize
\begin{eqnarray*}
U_{3}^{\prime } &=&\exp \left[ \frac{i\pi }{4\sqrt{2}}\left( -3E_{CD}-\frac{2}{3}\left( E_{AB}+E_{AC}+E_{BC}\right) \right. \right. \\
&&\left. \left. \;\;\;\;\;\;\;\;\;+J_{3}^{\prime }E_{AB}E_{CD}+J_{3}^{\prime
\prime }\left( E_{AC}E_{BD}+E_{AD}E_{BC}\right) \right) \right] \\
&=&\left( 
\begin{array}{cccccccccccccc}
\delta &  &  &  &  &  &  &  &  &  &  &  &  &  \\ 
& \delta &  &  &  &  &  &  &  &  &  &  &  &  \\ 
&  & \epsilon &  &  &  &  &  &  &  &  &  &  &  \\ 
&  &  & \epsilon &  &  &  &  &  &  &  &  &  &  \\ 
&  &  &  & \zeta &  &  &  &  &  &  &  &  &  \\ 
&  &  &  &  & \eta &  &  &  & \rho &  &  &  &  \\ 
&  &  &  &  &  & \zeta &  &  &  &  &  &  &  \\ 
&  &  &  &  &  &  & \eta &  &  &  &  &  & \rho \\ 
&  &  &  &  &  &  &  & \zeta &  &  &  &  &  \\ 
&  &  &  &  & \rho &  &  &  & \eta ^{*} &  &  &  &  \\ 
&  &  &  &  &  &  &  &  &  & \eta ^{*} &  & \rho &  \\ 
&  &  &  &  &  &  &  &  &  &  & e^{i\pi \frac{\left( -5+J_{3}^{\prime
}+2J_{3}^{\prime \prime }\right) }{4\sqrt{2}}} &  &  \\ 
&  &  &  &  &  &  &  &  &  & \rho &  & \eta &  \\ 
&  &  &  &  &  &  & \rho &  &  &  &  &  & \eta ^{*}
\end{array}
\right) ,
\end{eqnarray*}
\normalsize
where 
\begin{eqnarray*}
\nu &=&\sqrt{3(J_{k}^{\prime })^{2}+3(J_{k}^{\prime \prime
})^{2}-6J_{k}^{\prime }J_{k}^{\prime \prime }-16J_{k}^{\prime
}+16J_{k}^{\prime \prime }+24} \\
\delta &=&e^{i\pi \frac{1}{4\sqrt{2}}\left( J_{k}^{\prime }+2J_{k}^{\prime
\prime }+3\right) } \\
\epsilon &=&e^{i\pi \frac{1}{4\sqrt{2}}\left( J_{k}^{\prime }+2J_{k}^{\prime
\prime }+3\right) } \\
\zeta &=&e^{-i\pi \frac{1}{4\sqrt{2}}\left( J_{k}^{\prime }+3\right) } \\
\eta &=&e^{-i\pi \frac{1}{4\sqrt{2}}}\left[ \cos \left( \frac{\pi \nu }{4\sqrt{6}}\right) -\frac{1}{\sqrt{3}\nu }(J_{k}^{\prime }-J_{k}^{\prime
\prime })\sin \left( \frac{\pi \nu }{4\sqrt{6}}\right) \right] \\
\rho &=&\frac{\sqrt{2}(3-J_{k}^{\prime }+J_{k}^{\prime \prime })}{\sqrt{3}\nu }e^{-\frac{i\pi }{4\sqrt{2}}\left[ 1+J_{k}^{\prime \prime }-\frac{\nu }{\sqrt{3}}\right] }\left[ 1-e^{\frac{i\pi \nu }{2\sqrt{6}}}\right] ,
\end{eqnarray*}
and $k=2$ ($3$) for $U_{2}^{\prime }$ ($U_{3}^{\prime }$). It turns out that
these two gates can act \textquotedblleft classically\textquotedblright \quad only
when $\eta =0$, which leads to a transcendental equation relating $J_{k}^{\prime }$ and $J_{k}^{\prime \prime }$ ($J_{k}^{\prime }$ = $J_{k}^{\prime \prime }$ is one set of solutions).

\subsubsection{The $U_{1}$ Gate}

The $U_{1}$ gate is qualitatively different from all the previous gates. It
involves an interaction between five dots and two code blocks. In this case,
the effect of the four-body interactions is generally to strongly interfere
with the action of $U_{1}$. Therefore we must carefully reexamine this gate
and consider whether it can be made compatible with the four-body effect.
The generator of $U_{1}$ is 
\begin{equation*}
H_{1}=E_{DE}+\frac{1}{2}\sum_{A=i<j}^{D}E_{ij},
\end{equation*}%
and the modified generator, in the presence of four-body interactions, is in
general 
\begin{equation*}
H_{1}^{\prime }=H_{1}+\sum_{A=i<j<k<l}^{E}J_{ij;kl}^{\prime }E_{ij}E_{kl}.
\end{equation*}
Therefore the new gate will have the form 
\begin{equation*}
U_{1}^{\prime }=\exp (\frac{i\pi }{\sqrt{3}}H_{1}^{\prime }).
\end{equation*}
There are symmetry relations between the constants $J_{ij;kl}^{\prime }$:\
the magnitudes of the exchange interactions in $H_{1}$ imply an equivalence
between spins $A,B,C$, but spins $D$ and $E$ are distinct. Thus there will
be four such constants, corresponding to the following sets of inequivalent
two-body pairings:

\begin{itemize}
\item one of spins $\{A,B,C\}$ coupled with spin $D$, but without spin $E$:\ 
$J_{a}^{\prime }\equiv J_{AB;CD}^{\prime }=J_{AC;BD}^{\prime
}=J_{BC;AD}^{\prime }$;

\item one of spins $\{A,B,C\}$ coupled with spin $E$, but without spin $D$:\ 
$J_{b}^{\prime }\equiv J_{AB;CE}^{\prime }=J_{AC;BE}^{\prime
}=J_{BC;AE}^{\prime }$;

\item spin $D$ coupled with spin $E$:\ $J_{c}^{\prime }\equiv
J_{AB;DE}^{\prime }=J_{AC;DE}^{\prime }=J_{BC;DE}^{\prime }$;

\item one of spins $\{A,B,C\}$ coupled with spin $D$, one with spin $E$: $J_{d}^{\prime }\equiv J_{AD;BE}^{\prime }=J_{AD;CE}^{\prime
}=J_{BD;AE}^{\prime }=J_{BD;CE}^{\prime }=J_{CD;AE}^{\prime
}=J_{CD;BE}^{\prime }$.
\end{itemize}

Thus $U_{1}^{\prime }$ can be written as 
\begin{widetext}
\begin{eqnarray}
U_{1}^{\prime } &=&\exp \left[ \frac{i\pi }{\sqrt{3}}\left( E_{DE}+\frac{1}{2}\sum_{A=i<j}^{D}E_{ij}+J_{a}^{\prime }\left( E_{AB}E_{CD}\right. \right.
\right.  \notag \\
&&\;\;\;\;\;\left. \left. \left. +E_{AC}E_{BD}+E_{AD}E_{BC}\right) \right.
\right.  \notag \\
&&\left. \left. +J_{b}^{\prime }\left(
E_{AB}E_{CE}+E_{AC}E_{BE}+E_{AE}E_{BC}\right) \right. \right.  \notag \\
&&\left. \left. +J_{c}^{\prime }\left(
E_{AB}E_{DE}+E_{AC}E_{DE}+E_{BC}E_{DE}\right) \right. \right.  \notag \\
&&\left. \left. +J_{d}^{\prime }\left(
E_{AD}E_{BE}+E_{AD}E_{CE}+E_{AE}E_{BD}\right. \right. \right.  \notag \\
&&\;\;\;\;\;\left. \left. \left.
+E_{AE}E_{CD}+E_{BD}E_{CE}+E_{BE}E_{CD}\right) \right) \right]
\label{eq:U1'} \\
&=&\left( 
\begin{array}{cccccccccccccc}
\chi _{+} &  &  &  & \lambda &  &  &  &  &  &  &  &  &  \\ 
& \chi _{+} &  &  &  &  & \lambda &  &  &  &  &  &  &  \\ 
&  & \chi _{+} &  &  & \lambda &  &  &  &  &  &  &  &  \\ 
&  &  & \chi _{+} &  &  &  & \lambda &  &  &  &  &  &  \\ 
\lambda &  &  &  & \chi _{-} &  &  &  &  &  &  &  &  &  \\ 
&  & \lambda &  &  & \chi _{-} &  &  &  &  &  &  &  &  \\ 
& \lambda &  &  &  &  & \chi _{-} &  &  &  &  &  &  &  \\ 
&  &  & \lambda &  &  &  & \chi _{-} &  &  &  &  &  &  \\ 
&  &  &  &  &  &  &  & \xi &  &  &  &  &  \\ 
&  &  &  &  &  &  &  &  & \theta &  &  &  &  \\ 
&  &  &  &  &  &  &  &  &  & \tau _{-} & \mu &  &  \\ 
&  &  &  &  &  &  &  &  &  & \mu & \tau _{+} &  &  \\ 
&  &  &  &  &  &  &  &  &  &  &  & \xi &  \\ 
&  &  &  &  &  &  &  &  &  &  &  &  & \theta%
\end{array}
\right) ,  \notag
\end{eqnarray}
\end{widetext}
where 
\begin{widetext}
\begin{eqnarray}
\label{eq:kappa}
x &=& \left(9\left( 1+4{J_{b}^{\prime }}-4{J_{d}^{\prime }}\right)
+48\left( {{J_{a}^{\prime }}}^{2}+{{J_{b}^{\prime }}}^{2}-{J_{b}^{\prime }}
{J_{d}^{\prime }}+{{J_{d}^{\prime }}}^{2}-{J_{a}^{\prime }}\left( {
J_{b}^{\prime }}+{J_{d}^{\prime }}\right) \right) \right)^{1/2}  \\
y &=& \left(3+8{{J_{a}^{\prime }}}^{2}+8{{J_{b}^{\prime }}}^{2}+9{
J_{c}^{\prime }-J_{b}^{\prime }}\left( 9+6{J_{c}^{\prime }}-4{
J_{d}^{\prime }}\right) +2{\left( 3{J_{c}^{\prime }}-2{J_{d}^{\prime }}
  \right) }^{2} \right. \notag \\
&& \left. -2{J_{a}^{\prime }}\left( -3+7{J_{b}^{\prime }}+3{
J_{c}^{\prime }}-2{J_{d}^{\prime }}\right) -6{J_{d}^{\prime }} \right)^{1/2}  \notag
\\
{\chi }_{\pm } &=&\left( \cos \left( \pi x/6\right) \pm 2i\sqrt{3}
\frac{\left( 2{J_{a}^{\prime }}-{J_{b}^{\prime }}-{J_{d}^{\prime }}\right) 
}{x}\sin \left( \pi x/6\right) \right) e^{i\pi \frac{\sqrt{3}}{6}
\left( {{1+2{J_{a}^{\prime }}+2{J_{b}^{\prime }}+2{J_{d}^{\prime }}}}
\right) } \notag \\
{\lambda } &=&\frac{3i\left( 1+2{J_{b}^{\prime }}-2{J_{d}^{\prime }}
\right) }{x}\sin \left( \pi x/6\right) e^{i\pi \frac{\sqrt{3}}{6}
\left( {{1+2{J_{a}^{\prime }}+2{J_{b}^{\prime }}+2{J_{d}^{\prime }}}}
\right) }  \notag \\
{\xi } &=&e^{i\pi \frac{1}{\sqrt{3}}\left( 2-{J_{a}^{\prime }}-{
J_{b}^{\prime }}+2{J_{d}^{\prime }}\right) }  \notag \\
{\theta } &=&\xi e^{i\pi \sqrt{3}{J_{c}^{\prime }}}  \notag \\
{\tau }_{\pm }{} &=&\left( \cos (\frac{\pi y}{\sqrt{6}})\pm \frac{
i\left( {{3+8{J_{a}^{\prime }}-7{J_{b}^{\prime }}-3{J_{c}^{\prime }}
+2{J_{d}^{\prime }}}}\right) }{2\sqrt{2}y}\sin (\frac{\pi y}{\sqrt{6}}
)\right) e^{i\pi \frac{1}{\sqrt{3}}\left( 2+{J_{a}^{\prime }}+{
J_{b}^{\prime }}-2{J_{d}^{\prime }}\right) }  \notag \\
{\mu } &=&\frac{\sqrt{15}\left( {J_{b}^{\prime }}-3{J_{c}^{\prime }}+2
{J_{d}^{\prime }-1}\right) }{2\sqrt{2}iy}\sin (\frac{\pi y}{\sqrt{6}}
)e^{i\pi \frac{1}{\sqrt{3}}\left( 2+{J_{a}^{\prime }}+{J_{b}^{\prime }}-2
{J_{d}^{\prime }}\right) }.  \notag
\end{eqnarray}
\end{widetext}
\newpage
In comparison, Bacon's $U_{1}$ gate has the form 
\begin{equation}
U_{1}=\left( 
\begin{array}{cccccccccccccc}
&  &  &  & {\Omega } &  &  &  &  &  &  &  &  &  \\ 
&  &  &  &  &  & {\Omega } &  &  &  &  &  &  &  \\ 
&  &  &  &  & {\Omega } &  &  &  &  &  &  &  &  \\ 
&  &  &  &  &  &  & {\Omega } &  &  &  &  &  &  \\ 
{\Omega } &  &  &  &  &  &  &  &  &  &  &  &  &  \\ 
&  & {\Omega } &  &  &  &  &  &  &  &  &  &  &  \\ 
& {\Omega } &  &  &  &  &  &  &  &  &  &  &  &  \\ 
&  &  & {\Omega } &  &  &  &  &  &  &  &  &  &  \\ 
&  &  &  &  &  &  &  & \Xi &  &  &  &  &  \\ 
&  &  &  &  &  &  &  &  & \Xi &  &  &  &  \\ 
&  &  &  &  &  &  &  &  &  & \Gamma & \Delta &  &  \\ 
&  &  &  &  &  &  &  &  &  & \Delta & \Theta &  &  \\ 
&  &  &  &  &  &  &  &  &  &  &  & \Xi &  \\ 
&  &  &  &  &  &  &  &  &  &  &  &  & \Xi
\end{array}
\right) ,  \label{eq:U1}
\end{equation}
where 
\begin{eqnarray*}
{\Omega } &=&ie^{i\frac{\pi }{2\sqrt{3}}} \\
\Xi &=&e^{i\frac{2\pi }{\sqrt{3}}} \\
{\Delta } &=&\frac{i}{2}\sqrt{\frac{5}{2}}e^{i\frac{2\pi }{\sqrt{3}}}\sin (\frac{\pi }{\sqrt{2}}) \\
{\Theta } &=&e^{i\frac{2\pi }{\sqrt{3}}}\left( \cos (\frac{\pi }{\sqrt{2}})+i\sqrt{\frac{3}{8}}\sin (\frac{\pi }{\sqrt{2}})\right) \\
{\Gamma } &=&e^{i\frac{2\pi }{\sqrt{3}}}\left( \cos (\frac{\pi }{\sqrt{2}})-i\sqrt{\frac{3}{8}}\sin (\frac{\pi }{\sqrt{2}})\right) ,
\end{eqnarray*}
and it may be verified that $U_{1}^{\prime }$ reduces to $U_{1}$ in the
limit that $J_{a}^{\prime },J_{b}^{\prime },J_{c}^{\prime },J_{d}^{\prime
}\rightarrow 0$. The $U_{1}^{\prime }$ gate, like the $U_{1}$ gate, is
applied at the beginning of the controlled-phase gate sequence, and hence
acts on the computational basis states. By comparing the explicit matrix
representations (\ref{eq:U1'}) and (\ref{eq:U1}) it is clear that the
crucial difference between $U_{1}^{\prime }$ and $U_{1}$ is the appearance
of the $\chi _{+}$ terms on the diagonal of $U_{1}^{\prime }$ (the
difference between $\Omega $ and $\lambda $ is irrelevant:\ it translates
into a global phase). In order for $U_{1}^{\prime }$\ to act like $U_{1}$,
i.e., in order for it not to prepare a superposition of code states and the
first four non-code states, $\chi _{+}$\ must vanish. Consulting the
expression (\ref{eq:kappa}) for $\chi _{+}$, it is evident that this leads
to a complicated transcendental equation relating the constants $J_{a}^{\prime },J_{b}^{\prime },J_{d}^{\prime }$ ~(but not involving $J_{c}^{\prime }$). A numerical solution of the condition $\chi _{+}=0$ leads
to the result that the constants $J_{a}^{\prime },J_{b}^{\prime
},J_{d}^{\prime }$ can take on an infinite set of rationally related values.
Upon setting the ratio $J_{b}^{\prime }/J_{d}^{\prime }$ to any rational
number (except $1$), there is a corresponding rational value of $J_{a}^{\prime }$.

\subsubsection{Summary of Conditions}

Summing up our findings, we have the following sufficient set of conditions
for the modified gate sequence 
\begin{widetext}
\begin{equation*}
CP^{\prime }=U_{1}^{\prime \dagger }\left[ U_{2}^{\prime \dagger
}U_{3}^{\prime \dagger }\right] U_{5}^{\prime \dagger }U_{6}^{\prime
}(J_{B}^{\prime })U_{5}^{\prime }\left[ U_{3}^{\prime }(J_{3}^{\prime
})U_{2}^{\prime }(J_{2}^{\prime })\right] U_{1}^{\prime }(J_{a}^{\prime
},J_{b}^{\prime },J_{c}^{\prime },J_{d}^{\prime })
\end{equation*}
\end{widetext}
to work as a controlled-phase gate in the presence of
four-body interactions:

\begin{enumerate}
\item The constant $J_{c}^{\prime }$ can take on an arbitrary value.

\item The constants $J_{2}^{\prime }$ and $J_{2}^{\prime \prime }$ must be
chosen to satisfy the transcendental equation $\eta =0$.

\item The constants $J_{3}^{\prime }$ and $J_{3}^{\prime \prime }$ must be
chosen to satisfy the transcendental equation $\eta =0$.

\item The constant $J_{5}^{\prime }$ must either be zero or chosen such that 
$\Lambda $ is an even integer (i.e. $\sqrt{\scriptstyle\frac{4}{3}\displaystyle(J_{5}^{\prime })^{2}-2J_{5}^{\prime }+1}$ is an even integer).

\item The constant $J_{B}^{\prime }$ must be an integer.

\item The constants $J_{a}^{\prime },J_{b}^{\prime },J_{d}^{\prime }$ can
take on an infinite set of rationally related values, where the ratio of any
pair (e.g., $J_{b}^{\prime }/J_{d}^{\prime }$)\ can be chosen completely
arbitrarily, and the value of the third constant is determined by this
choice.
\end{enumerate}

The most restrictive of these conditions is that $J_{B}^{\prime }$ must be
an integer. However, note that since the gates are applied sequentially,
this condition need only be satisfied during the application of the $U_{6}^{\prime }$ gate, and it is plausible from the earlier sections of this
paper that corresponding Heisenberg exchange constants can be found. When
these conditions are satisfied it is indeed the case that $CP^{\prime
}=(-1,1,1,1)$ on the code space.

\subsection{Dimensionality of Parameter Spaces Required by Two-Body and
Four-Body Couplings}

We caution that, although the encoding procedure described above has been
shown mathematically to remove the effect of the four-body couplings, the
experimental construction of a suitable apparatus using real quantum dots is
another matter, as the following heuristic calculation suggests.

Our modified gates imply the following constraints on the coupling
coefficients:

\underline{$U_{5}^{\prime }$ gate}

\begin{enumerate}
\item[(a)] $K_{2}[FG]=\scriptstyle\frac{1}{2}\displaystyle K_{2}[GH]$;

\item[(b)] Either $K_{2}[FH]=0$, or $\Lambda (K_{2}[FH])=2n$, where $n$ must
have an integer value;
\end{enumerate}

\underline{$U_{B}^{\prime }$ gate}

\begin{enumerate}
\item[(c)] $K_{2}[ij]$ is the same for all pairs within $\{A,B,C,D\}$;

\item[(d)] $K_{4}[ABCD]=K_{4}[ACBD]=K_{4}[ADBC]$;

\item[(e)] $K_{4}[ABCD]=2mK_{2}[AB]$, where $m$ may have any integer value;
\end{enumerate}

\underline{$U_{2}^{\prime }$ gate}

\begin{enumerate}
\item[(f)] $K_{2}[FG]=K_{2}[FH]=K_{2}[GH]$;

\item[(g)] $K_{2}[EF]=\scriptstyle\frac{9}{2}\displaystyle K_{2}[GH]$;

\item[(h)] $K_{4}[EGFH]=K_{4}[EHFG]$;

\item[(i)] Either $K_{4}[EFGH]=K_{4}[EGFH]$, or $K_{4}[EFGH]$ and $K_{4}[EGFH]$ satisfy the transcendental equation $\eta
(K_{4}[EFGH],K_{4}[EGFH])=0$;
\end{enumerate}

\underline{$U_{3}^{\prime }$ gate}

\begin{enumerate}
\item[(j)] $K_{2}[AB]=K_{2}[AC]=K_{2}[BC]$;

\item[(k)] $K_{2}[CD]=\scriptstyle\frac{9}{2}\displaystyle K_{2}[AB]$;

\item[(l)] $K_{4}[ACBD]=K_{4}[ADBC]$;

\item[(m)] Either $K_{4}[ABCD]=K_{4}[ACBD]$, or $K_{4}[ABCD]$ and $K_{4}[ACBD]$ satisfy the transcendental equation $\eta
(K_{4}[ABCD],K_{4}[ACBD])=0$;
\end{enumerate}

\underline{$U_{1}^{\prime }$ gate}

\begin{enumerate}
\item[(n)] $K_{2}[ij]$ is the same for all pairs within $\{A,B,C,D\}$;

\item[(o)] $K_{2}[DE]=2K_{2}[AB]$;

\item[(p)] $K_{4}[ABCD]=K_{4}[ACBD]=K_{4}[ADBC]$;

\item[(q)] $K_{4}[ABCE]=K_{4}[ACBE]=K_{4}[AEBC]$;

\item[(r)] $K_{4}[ADBE]=K_{4}[AEBD]$;

\item[(s)] $K_{4}[ADCE]=K_{4}[AECD]$;

\item[(t)] $K_{4}[BDCE]=K_{4}[BECD]$;

\item[(u)] $K_{4}[ADBE]=K_{4}[ADCE]=K_{4}[BDCE]$;

\item[(v)] $K_{4}[ABDE]$ is a single-valued function of $K_{4}[ADBE]$;

\item[(w)] $K_{4}[BCDE]$ is a single-valued function of $K_{4}[ADCE]$;

\item[(x)] $K_{4}[ACDE]$ is a single-valued function of $K_{4}[BDCE]$.
\end{enumerate}

Since the coupling coefficients must in general vary with time in order to
satisfy all of these constraints (for example, $K_{2}[CD]$ and $K_{2}[AC]$
would be equal during the operation of $U_{B}^{\prime }$, but unequal during 
$U_{3}^{\prime }$), we also assume that particular constraints need to be
concurrently satisfied only when they arise from the same gate.

First, by the same reasoning used to derive Eqs.~(\ref{HspinJ_3}) and (\ref{HspinJ_4}), we note that a four-dot Hamiltonian for the geometry of $\{A,B,C,E\}$ contains a constant term and 9 independent coupling
coefficients. If these 9 coefficients take on a given set of values and we
wish to adjust them to meet constraints such as those listed above, we would
need 9 additional degrees of freedom in the system. We make the conservative
assumption, however, that one two-body coefficient and one four-body
coefficient can be left unaltered and the others adjusted to correspond to
them, which means that only 7 additional parameters are required. Similarly,
for the subset $\{A,B,D,E\}$ ($\{A,C,D,E\}$, $\{B,C,D,E\}$, $\{A,B,C,D\}$),
there are 9 (6, 7, 5) independent coupling coefficients, for which we
require 7 (4, 5, 3) tunable parameters if a given set of constraints are to
be satisfied. Of course, we will count one more degree of freedom whenever a
constraint includes relationships between the two-body and four-body
energies.

Now suppose that we designate one \textquotedblleft base\textquotedblright \quad choice of $\{x_{b},x_{c},x_{v}\}$, such that within each of the two squares,
all the quantities $K_{2}[ij]$ are equal, all the quantities $K_{4}[ijkl]$
are equal, and $K_{4}[ijkl]=2K_{2}[ij]$. That arrangement can simultaneously
satisfy constraints (b), (c), (d), (e), (f), (h), (i), (j), (l), (m), (n),
and (p), provided that the value of $K_{2}[FH]$ is chosen appropriately.
From this potential, we would need to make one change within $\{E,F,G,H\}$
to reach condition (a) or condition (g), or one change within $\{5,6,7,8\}$
to obtain (k) or (o). The couplings of $\{A,B,C,E\}$ must be adjusted to
match (q) while still satisfying (n), (o), and (p), which requires 6
additional degrees of freedom as explained in the previous paragraph.
Similarly, (v) ((w), (x)) and (r) ((s), (t)) together imply particular
adjustments to the four-body couplings in $\{A,B,D,E\}$ ($\{A,C,D,E\}$, $\{B,C,D,E\}$), which requires 6 (3,4) new parameters. (The single-valued
function in question is the same for all three cases, so $K_{4}[ABDE]$ ends
up equalling $K_{4}[ACDE]$ and $K_{4}[BCDE]$.) Finally, we need two more
degrees of freedom available somewhere in order to meet constraint (u), for
a grand total of 28 degrees of freedom.

To put the size of this number into perspective, we will also count the
independently tuned energies necessary to meet the conditions on EQC using pairwise couplings alone. By choosing a suitable
combination of $\{x_{b},x_{c},x_{v}\}$ for an entire eight-spin system, we
could satisfy (b), (c), (f), (j), and (n) at the same time; one more degree
of freedom would be necessary to also satisfy (o). Starting from such a
system, we could presumably satisfy (a) or (g) by adjusting one parameter
within $\{E,F,G,H\}$, or satisfy (k) by adjusting one parameter within $\{A,B,C,D\}$. Hence we estimate that 7 degrees of freedom are required for
the purely Heisenberg Hamiltonian used in Ref. \cite{Bacon:thesis}. We see
that, even if one presupposes the ability to create and position many
identical qubits of the form (\ref{eq:0L}), (\ref{eq:1L}) (3 free
parameters), accounting correctly for two-body and four-body coupling is
still a great deal more demanding than two-body coupling alone. It
is this experimental challenge that must be weighed against the increased
length (and hence vulnerability to decoherence) of pulse sequences employing
only two-body couplings \cite{Hsieh:04}.

\section{Summary and Conclusions}

Earlier work \cite{MizelLidar,MizelLidarB} showed that in highly
symmetrical geometries, the interaction between three and four mutually
interacting electrons confined in parabolic potentials contains many-body
terms, which in the case of four electrons qualitatively modify the usual
Heisenberg interaction. In this work we have improved upon these early
results by considering realistic, linear and square geometries, and by
utilizing Gaussian confining potentials. Specifically, we have shown in a
Heitler-London calculation that in the case of four mutually interacting
electrons, in both the linear and square geometries, the system's
Hamiltonian contains four-body exchange terms which may be of comparable
strength to the Heisenberg exchange interactions. This can have important
implications for quantum information processing using coupled quantum dots. We have
considered, in particular, the implications for the quantum computing using
logical qubits encoded into decoherence-free subspaces of four electrons per qubit. We
showed that previously designed conditional quantum logic gates between
these encoded qubits must be modified, in order to account for the four-body
terms that alter the (previously assumed) Heisenberg interaction, when four
or more electrons are coupled simultaneously. This requires the ability to tune, to a certain extent, the four-body exchange
constants. It is worth noting, however, that there are alternatives
to this method of implementing encoded conditional logic gates, which may be
less demanding. In particular, it is worth exploring the possibility of
completing the set of universal encoded quantum logic gates by supplementing
single-qubit gates (where, as we have shown, four-body effects are
harmless) with measurements and teleportation, as in linear optics quantum
computing \cite{KLM-Nature}. This will be a subject for future research.

\section*{Acknowledgments}

A.M. and R.W. acknowledge the support of the Packard Foundation. D.A.L. acknowledges support under the DARPA-QuIST program (managed by AFOSR under agreement No. F49620-01-1-0468), and the Sloan Foundation. R.W. is grateful to Dr. Rusko Ruskov for constructive discussion.

\newpage

{\normalsize 
\begin{widetext}
\begin{figure}[tbp]
  \centering
  \includegraphics[height=20cm,angle=270]{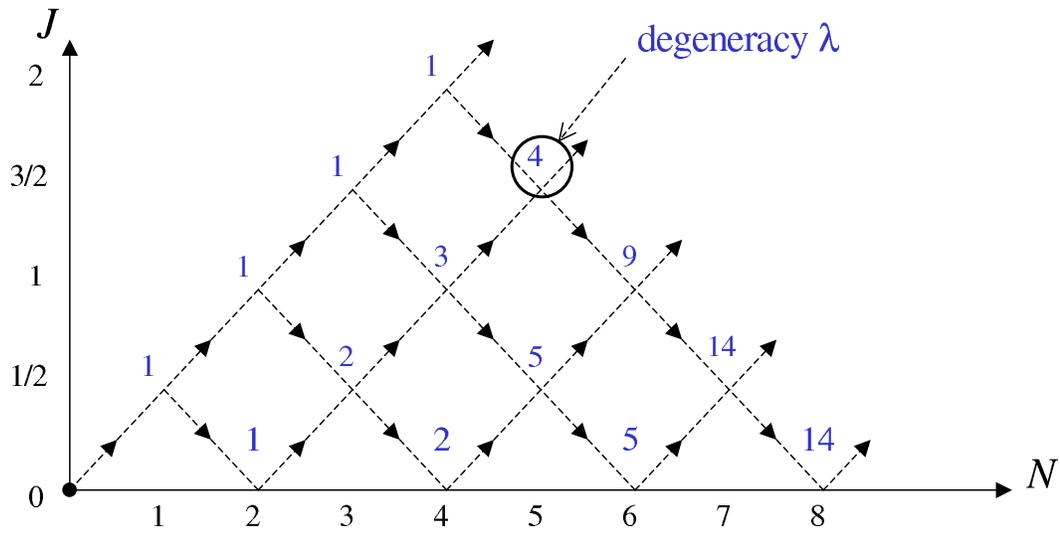}
  \vspace{-5cm}
\caption{(Color online) Partitioning of the Hilbert space of $N$ spin-$1/2$ particles into
DF subspaces (nodes of the graph). The integer above each node represents the
number of paths leading from the origin to that node.}
\label{fig:DFS}
\end{figure}
\end{widetext}
}


\begin{thebibliography}{57}
\expandafter\ifx\csname natexlab\endcsname\relax\def\natexlab#1{#1}\fi
\expandafter\ifx\csname bibnamefont\endcsname\relax
  \def\bibnamefont#1{#1}\fi
\expandafter\ifx\csname bibfnamefont\endcsname\relax
  \def\bibfnamefont#1{#1}\fi
\expandafter\ifx\csname citenamefont\endcsname\relax
  \def\citenamefont#1{#1}\fi
\expandafter\ifx\csname url\endcsname\relax
  \def\url#1{\texttt{#1}}\fi
\expandafter\ifx\csname urlprefix\endcsname\relax\def\urlprefix{URL }\fi
\providecommand{\bibinfo}[2]{#2}
\providecommand{\eprint}[2][]{\url{#2}}

\bibitem{LossDiVincenzo98} 
\bibinfo{author}{\bibnamefont{{D. Loss and D.P.
DiVincenzo}}}, \bibinfo{journal}{Phys. Rev. A} \textbf{\bibinfo{volume}{57}}, 
\bibinfo{pages}{120} (\bibinfo{year}{1998}).

\bibitem{MAN/ILC} 
\bibinfo{author}{
\bibnamefont{{M.A. Nielsen and I.L. Chuang}}}, 
\bibinfo{title}{Quantum
Computation and Quantum Information} (
\bibinfo{publisher}{Cambridge
University Press}, \bibinfo{address}{Cambridge, U.K.}, \bibinfo{year}{2000}) 
\bibinfo{pages}{191-193}.

\bibitem{Preskill} \bibinfo{author}{\bibnamefont{{J. Preskill}}}, 
\bibinfo{journal}{Proc. Roy. Soc. Lond. A} \textbf{\bibinfo{volume}{454}}, 
\bibinfo{pages}{385-410} (\bibinfo{year}{1998}).

\bibitem[{R. Raussendorf and H.J. Briegel}(2001)]{Raussendorf:01} 
\bibinfo{author}{\bibnamefont{{R. Raussendorf and H.J. Briegel}}}, 
\bibinfo{journal}{Phys. Rev. Lett.} \textbf{\bibinfo{volume}{86}}, 
\bibinfo{pages}{5188} (\bibinfo{year}{2001}).

\bibitem[{D. Bacon, J. Kempe, D.A. Lidar and K.B. Whaley}(2000)]{Bacon:99a} 
\bibinfo{author}{\bibnamefont{{D. Bacon, J. Kempe, D.A. Lidar and K.B.
  Whaley}}}, \bibinfo{journal}{Phys. Rev. Lett.} \textbf{\bibinfo{volume}{85}}, 
\bibinfo{pages}{1758} (\bibinfo{year}{2000}).

\bibitem[{J. Kempe, D. Bacon, D.A. Lidar, and K.B. Whaley}(2001)]{Kempe:00} 
\bibinfo{author}{\bibnamefont{{J. Kempe, D. Bacon, D.A. Lidar, and K.B.
  Whaley}}}, \bibinfo{journal}{Phys. Rev. A} \textbf{\bibinfo{volume}{63}}, 
\bibinfo{pages}{042307} (\bibinfo{year}{2001}).

\bibitem[{D.P. DiVincenzo, D. Bacon, J. Kempe, G. Burkard, and K.B. Whaley}
(2000)]{DiVincenzo:00a} 
\bibinfo{author}{\bibnamefont{{D.P. DiVincenzo, D. Bacon, J. Kempe, G. Burkard,
  and K.B. Whaley}}}, \bibinfo{journal}{Nature} \textbf{\bibinfo{volume}{408}}, 
\bibinfo{pages}{339} (\bibinfo{year}{2000}).

\bibitem[{D. Bacon, J. Kempe, D.P. DiVincenzo, D.A. Lidar, and K.B. Whaley}
(2001)]{Bacon:Sydney} 
\bibinfo{author}{\bibnamefont{{D. Bacon, J. Kempe, D.P. DiVincenzo, D.A. Lidar,
  and K.B. Whaley}}}, in 
\bibinfo{booktitle}{Proceedings of the 1st
  International Conference on Experimental Implementations of Quantum
  Computation, Sydney, Australia}, edited by \bibinfo{editor}{
\bibfnamefont{R.}~\bibnamefont{Clark}} (\bibinfo{publisher}{Rinton}, 
\bibinfo{address}{Princeton, NJ}, \bibinfo{year}{2001}), p. 
\bibinfo{pages}{257}.

\bibitem[{D.A. Lidar and L.-A. Wu}(2002)]{LidarWu:01} \bibinfo{author}{
\bibnamefont{{D.A. Lidar and L.-A. Wu}}}, \bibinfo{journal}{Phys. Rev. Lett.}
\textbf{\bibinfo{volume}{88}}, \bibinfo{pages}{017905} (\bibinfo{year}{2002}).

\bibitem[{D. Bacon, K.R. Brown, K.B. Whaley}(2001)]{Bacon:01} 
\bibinfo{author}{\bibnamefont{{D. Bacon, K.R. Brown, K.B. Whaley}}}, 
\bibinfo{journal}{Phys. Rev. Lett.} \textbf{\bibinfo{volume}{87}}, 
\bibinfo{pages}{247902} (\bibinfo{year}{2001}).

\bibitem[{E. Farhi, J. Goldstone, S. Gutmann, J. Lapan, A. Lundgren, D. Preda
}(2001)]{Farhi:01} 
\bibinfo{author}{\bibnamefont{{E. Farhi, J. Goldstone, S.
Gutmann, J. Lapan, A. Lundgren, D. Preda}}}, \bibinfo{journal}{Science} 
\textbf{\bibinfo{volume}{292}}, \bibinfo{pages}{472} (\bibinfo{year}{2001}).

\bibitem[{P.W. Shor}(1996)]{Shor:96} 
\bibinfo{author}{\bibnamefont{{P.W.
Shor}}}, in 
\bibinfo{booktitle}{{Proceedings of the 37th Symposium on Foundations
  of Computing}} (\bibinfo{publisher}{{IEEE Computer Society Press}}, 
\bibinfo{address}{{Los Alamitos, CA}}, \bibinfo{year}{1996}), p.~ 
\bibinfo{pages}{56}.

\bibitem[Gottesman(1997)]{Gottesman:97a} 
\bibinfo{author}{\bibfnamefont{D.}~
\bibnamefont{Gottesman}}, \bibinfo{journal}{Phys. Rev. A} 
\textbf{\bibinfo{volume}{57}}, \bibinfo{pages}{127} (\bibinfo{year}{1997}).

\bibitem[{A.M. Steane}(1999)]{Steane:99a} 
\bibinfo{author}{
\bibnamefont{{A.M. Steane}}}, \bibinfo{journal}{Nature} \textbf{\bibinfo{volume}{399}}
, \bibinfo{pages}{124} (\bibinfo{year}{1999}).

\bibitem[{D.A. Lidar, D. Bacon, J. Kempe, and K.B. Whaley}(2001)]{Lidar:00b} 
\bibinfo{author}{\bibnamefont{{D.A. Lidar, D. Bacon, J. Kempe, and K.B.
  Whaley}}}, \bibinfo{journal}{Phys. Rev. A} \textbf{\bibinfo{volume}{63}}, 
\bibinfo{pages}{022307} (\bibinfo{year}{2001}).

\bibitem[{M.H. Freedman}(2003)]{Freedman:02} 
\bibinfo{author}{
\bibnamefont{{M.H. Freedman}}}, \bibinfo{journal}{Commun. Math. Phys.} 
\textbf{\bibinfo{volume}{234}}, \bibinfo{pages}{129} (\bibinfo{year}{2003}).

\bibitem{MizelLidar} 
\bibinfo{author}{\bibnamefont{{A. Mizel and D.A. Lidar}}},
\bibinfo{journal}{Phys. Rev. Lett.}
\textbf{\bibinfo{volume}{92}}, 
\bibinfo{pages}{077903} (\bibinfo{year}{2004}).

\bibitem{MizelLidarB} 
\bibinfo{author}{\bibnamefont{{A. Mizel and D.A. Lidar}}},
\bibinfo{journal}{Phys. Rev. B}
\textbf{\bibinfo{volume}{70}}, 
\bibinfo{pages}{115310} (\bibinfo{year}{2004}).

\bibitem{3inaline_figure} 
\bibinfo{author}{\bibnamefont{{F.R. Waugh, M.J. Berry, C.H. Crouch, C. Livermore, D.J. Mar, R.M. Westervelt, K.L. Campman, and A.C. Gossard}}},
\bibinfo{journal}{Phys. Rev. B}
\textbf{\bibinfo{volume}{53}}, 
\bibinfo{pages}{1413} (\bibinfo{year}{1996}).

\bibitem[{G. Burkard, H.-A. Engel and D. Loss}(2000)]{Burkard:00} 
\bibinfo{author}{\bibnamefont{{G. Burkard, H.-A. Engel and D. Loss}}}, 
\bibinfo{journal}{Fortschr. Phys.} \textbf{\bibinfo{volume}{48}}, \bibinfo{pages}{965}
(\bibinfo{year}{2000}).

\bibitem{Hu:99} \bibinfo{author}{\bibnamefont{{X. Hu and  S. Das Sarma}}}, 
\bibinfo{journal}{Phys. Rev. A} \textbf{\bibinfo{volume}{61}}, \bibinfo{pages}{062301}
(\bibinfo{year}{2000}).

\bibitem[{V.W. Scarola, K. Park, and S. Das Sarma}(2004)]{Scarola:04} 
\bibinfo{author}{\bibnamefont{{V.W. Scarola, K. Park, and S. Das Sarma}}}, 
\bibinfo{journal}{Phys. Rev. Lett.} \textbf{\bibinfo{volume}{93}}, \bibinfo{pages}{120503} (\bibinfo{year}{2004}).

\bibitem[{V.W. Scarola and S. Das Sarma}(2005)]{Scarola:05} 
\bibinfo{author}{\bibnamefont{{V.W. Scarola and S. Das Sarma}}}, 
\bibinfo{journal}{Phys. Rev. A} \textbf{\bibinfo{volume}{71}}, \bibinfo{pages}{032340} (\bibinfo{year}{2005}).

\bibitem[Zanardi(1999)]{Zanardi:99a} \bibinfo{author}{\bibfnamefont{P.}~
\bibnamefont{Zanardi}}, \bibinfo{journal}{Phys. Rev. A} \textbf{\bibinfo{volume}{60}}
, \bibinfo{pages}{R729} (\bibinfo{year}{1999}).

\bibitem[{L.-A. Wu and D.A. Lidar}(2002)]{WuLidar:01b} \bibinfo{author}{
\bibnamefont{{L.-A. Wu and D.A. Lidar}}}, \bibinfo{journal}{Phys. Rev. Lett.}
\textbf{\bibinfo{volume}{88}}, \bibinfo{pages}{207902} (\bibinfo{year}{2002}).

\bibitem[{L.-A. Wu, M.S. Byrd, D.A. Lidar}(2002)]{WuByrdLidar:02} 
\bibinfo{author}{\bibnamefont{{L.-A. Wu, M.S. Byrd, D.A. Lidar}}}, 
\bibinfo{journal}{Phys. Rev. Lett.} \textbf{\bibinfo{volume}{89}}, 
\bibinfo{pages}{127901} (\bibinfo{year}{2002}).

\bibitem[{P. Zanardi and M. Rasetti}(1997)]{Zanardi:97c} \bibinfo{author}{
\bibnamefont{{P. Zanardi and M. Rasetti}}}, 
\bibinfo{journal}{Phys. Rev.
Lett.} \textbf{\bibinfo{volume}{79}}, \bibinfo{pages}{3306} (\bibinfo{year}{1997}).

\bibitem[{D.A. Lidar, D. Bacon, J. Kempe and K.B. Whaley}(2000)]
{Lidar:PRA00Exchange} 
\bibinfo{author}{\bibnamefont{{D.A. Lidar, D. Bacon, J. Kempe and K.B.
  Whaley}}}, \bibinfo{journal}{Phys. Rev. A} \textbf{\bibinfo{volume}{61}}, 
\bibinfo{pages}{052307} (\bibinfo{year}{2000}).

\bibitem[{D.A. Lidar, K.B. Whaley}(2003)]{Lidar:DFSreview}
\bibinfo{author}{\bibnamefont{{D.A. Lidar and K.B. Whaley}}}, in 
\bibinfo{booktitle}{{Proceedings of the 37th Symposium on Foundations of Computing}},
\bibinfo{editor}{F. Benatti and R. Floreanini}, eds.
(\bibinfo{publisher}{{Springer}}, 
\bibinfo{address}{{Berlin}}, \bibinfo{year}{2003}), p.~\bibinfo{pages}{83},
\eprint{quant-ph/0301032}.

\bibitem[{D.M. Bacon}(2001)]{Bacon:thesis} \bibinfo{author}{
\bibnamefont{{D.M. Bacon}}}, Ph.D. thesis, 
\bibinfo{school}{Univ. of
California, Berkeley} (\bibinfo{year}{2001}), \eprint{quant-ph/0305025}.

\bibitem[{M. Hsieh, J. Kempe, S. Myrgren, K.B. Whaley}(2004)]{Hsieh:04} 
\bibinfo{author}{
\bibnamefont{{M. Hsieh, J. Kempe, S. Myrgren, K.B. Whaley}}}, \bibinfo{journal}{Q. Inf. Processing} 
\textbf{\bibinfo{volume}{2}}, \bibinfo{pages}{289} (\bibinfo{year}{2004}).

\bibitem{HL} \bibinfo{author}{
\bibnamefont{{W. Heitler and F. London}}}, \bibinfo{journal}{Z. Physik} 
\textbf{\bibinfo{volume}{44}}, \bibinfo{pages}{455} (\bibinfo{year}{1927}).

\bibitem[{G. Burkard, D. Loss and D.P. DiVincenzo}(1999)]{Burkard:99} 
\bibinfo{author}{\bibnamefont{{G. Burkard, D. Loss and D.P. DiVincenzo}}}, 
\bibinfo{journal}{Phys. Rev. B} \textbf{\bibinfo{volume}{59}}, \bibinfo{pages}{2070} (\bibinfo{year}{1999}).

\bibitem[{X. Hu and S. Das Sarma}(2001)]{Hu:01b} \bibinfo{author}{
\bibnamefont{{X. Hu and S. Das Sarma}}}, \bibinfo{journal}{Phys. Rev. A} 
\textbf{\bibinfo{volume}{64}}, \bibinfo{pages}{042312} (\bibinfo{year}{2001}).

\bibitem[{J. Levy}(2002{a})]{Levy:01} 
\bibinfo{author}{\bibnamefont{{J.
Levy}}}, \bibinfo{journal}{Phys. Rev. Lett.} \textbf{\bibinfo{volume}{89}}, 
\bibinfo{pages}{147902} (\bibinfo{year}{2002}{\natexlab{a}}).

\bibitem[{J. Levy}(2002{b})]{Levy:02} 
\bibinfo{author}{\bibnamefont{{J.
Levy}}}, \bibinfo{journal}{Phys. Stat. Sol. B} \textbf{\bibinfo{volume}{233}}, 
\bibinfo{pages}{467} (\bibinfo{year}{2002}{\natexlab{a}}).

\bibitem[{R. de Sousa and S. Das Sarma}(2003)]{sousa:115322} 
\bibinfo{author}{\bibnamefont{{R. de Sousa and S. Das Sarma}}}, 
\bibinfo{journal}{Phys. Rev. B} \textbf{\bibinfo{volume}{68}}, \bibinfo{pages}{115322} (\bibinfo{year}{2003}).

\bibitem[{E. Yablonovitch, H.W. Jiang, H. Kosaka, H.D. Robinson, D.S. Rao, and T. Szkopek}(2003)]{Yablonovitch:03} 
\bibinfo{author}{
\bibnamefont{{E. Yablonovitch, H.W. Jiang, H. Kosaka, H.D. Robinson, D.S. Rao, and T. Szkopek}}}, \bibinfo{journal}{Proc. of the IEEE} 
\textbf{\bibinfo{volume}{91}}, \bibinfo{pages}{761} (\bibinfo{year}{2003}).

\bibitem[{A. Barenco, C.H. Bennett, R. Cleve, D.P. DiVincenzo, N. Margolus,
P. Shor, T. Sleator, J. Smolin and H. Weinfurter}(1995)]{Barenco:95a} 
\bibinfo{author}{\bibnamefont{{A. Barenco, C.H. Bennett, R. Cleve, D.P.
  DiVincenzo, N. Margolus, P. Shor, T. Sleator, J. Smolin and H. Weinfurter}}}
, \bibinfo{journal}{Phys. Rev. A} \textbf{\bibinfo{volume}{52}}, \bibinfo{pages}{3457}
(\bibinfo{year}{1995}).

\bibitem[{E. Knill, R. Laflamme, and G. J. Milburn}(2001)]{KLM-Nature} 
\bibinfo{author}{
\bibnamefont{{E. Knill, R. Laflamme, and G. J. Milburn}}}, \bibinfo{journal}{Nature} 
\bibinfo{volume}{409}, \textbf{\bibinfo{volume}{46}} (\bibinfo{year}{2001}).

\end{thebibliography}
\end{document}